%% file: ms.tex
\documentclass[twocolumn]{aastex63} 

\newcommand{\teff}{$T_{\rm{eff}}$}
\newcommand{\kms}{km~s$^{-1}$}
\usepackage{xcolor}	
\usepackage{graphicx}	
\usepackage{float}

\begin{document} 

\title{Visual Orbits of Spectroscopic Binaries with the CHARA Array. IV. \\
HD~61859, HD~89822, HD~109510, and HD~191692}

\author[0000-0002-9903-9911]{Kathryn V. Lester} 
\affil{NASA Ames Research Center, Moffett Field, CA 94035, USA}

\author[0000-0001-5415-9189]{Gail H. Schaefer} 
\affil{The CHARA Array of Georgia State University, Mount Wilson Observatory, Mount Wilson, CA 91023, USA}

\author[0000-0002-9413-3896]{Francis C. Fekel} 
\affil{Center of Excellence in Information Systems, Tennessee State University, Nashville, TN 37209, USA}

\author[0000-0001-8537-3583]{Douglas R. Gies} 
\affil{Center for High Angular Resolution Astronomy and Department of Physics \& Astronomy, Georgia State University, Atlanta, GA 30302, USA}

\author[0000-0002-9061-2865]{Todd J. Henry} 
\affil{RECONS Institute, Chambersburg, PA 17201, USA}

\author[0000-0003-0193-2187]{Wei-Chun Jao} 
\affil{Department of Physics \& Astronomy, Georgia State University, Atlanta, GA 30302, USA}

\author[0000-0003-1324-0495]{Leonardo A. Paredes} 
\affil{Department of Physics \& Astronomy, Georgia State University, Atlanta, GA 30302, USA}

\author[0000-0003-4568-2079]{Hodari-Sadiki Hubbard-James} 
\affil{Department of Physics \& Astronomy, Georgia State University, Atlanta, GA 30302, USA}

\author[0000-0001-9939-2830]{Christopher D. Farrington} 
\affil{The CHARA Array of Georgia State University, Mount Wilson Observatory, Mount Wilson, CA 91023, USA}

\author[0000-0003-1338-531X]{Kathryn D. Gordon} 
\affil{University of Tampa, Tampa, FL 33606, USA}

\author[0000-0001-9984-0891]{S. Drew Chojnowski} 
\affil{Department of Physics, Montana State University, Bozeman, MT 59717, USA}

\author[0000-0002-3380-3307]{John D. Monnier}  
\affil{Department of Astronomy, University of Michigan, Ann Arbor, MI 48109, USA}

\author[0000-0001-6017-8773]{Stefan Kraus} 
\affil{Astrophysics Group, Department of Physics \& Astronomy, University of Exeter, Exeter, EX4 4QL, UK}

\author[0000-0002-0493-4674]{Jean-Baptiste Le Bouquin} 
\affil{Univ. Grenoble Alpes, CNRS, IPAG, 38000, Grenoble, France}

\author[0000-0002-2208-6541]{Narsireddy Anugu} 
\affil{The CHARA Array of Georgia State University, Mount Wilson Observatory, Mount Wilson, CA 91023, USA}

\author[0000-0002-0114-7915]{Theo ten Brummelaar}   
\affil{The CHARA Array of Georgia State University, Mount Wilson Observatory, Mount Wilson, CA 91023, USA}

\author[0000-0001-9764-2357]{Claire L. Davies} 
\affil{Astrophysics Group, Department of Physics \& Astronomy, University of Exeter, Exeter, EX4 4QL, UK}

\author[0000-0002-3003-3183]{Tyler Gardner} 
\affil{Department of Astronomy, University of Michigan, Ann Arbor, MI 48109, USA}

\author[0000-0001-8837-7045]{Aaron Labdon} 
\affil{European Southern Observatory, Casilla 19001, Santiago 19, Chile}

\author[0000-0001-9745-5834]{Cyprien Lanthermann} 
\affil{The CHARA Array of Georgia State University, Mount Wilson Observatory, Mount Wilson, CA 91023, USA}

\author[0000-0001-5980-0246]{Benjamin R. Setterholm} 
\affil{Department of Astronomy, University of Michigan, Ann Arbor, MI 48109, USA}

\correspondingauthor{Kathryn Lester}
\email{kathryn.v.lester@nasa.gov}

% -----------------------------------------------------------------------------
\begin{abstract}

We present the visual orbits of four spectroscopic binary stars, HD~61859, HD~89822, HD~109510, and HD~191692, using long baseline interferometry with the CHARA Array. We also obtained new radial velocities from echelle spectra using the APO 3.5~m, CTIO 1.5~m, and Fairborn Observatory 2.0~m telescopes. 
By combining the astrometric and spectroscopic observations, we solve for the full, three-dimensional orbits and determine the stellar masses to $1-12$\% uncertainty and distances to $0.4-6$\% uncertainty. We then estimate the effective temperature and radius of each component star through Doppler tomography and spectral energy distribution analyses. We found masses of $1.4-3.5M_\odot$, radii of $1.5-4.7R_\odot$, and temperatures of $6400-10300$K. We then compare the observed stellar parameters to the predictions of the stellar evolution models, but found that only one of our systems fits well with the evolutionary models.

\end{abstract}

\keywords{binaries: spectroscopic, binaries: visual, stars: fundamental parameters}

% -----------------------------------------------------------------------------
\section{Introduction}
We continue our paper series of determining the visual orbits of spectroscopic binary stars with long baseline interferometry -- Paper I \citep{lester19a}, Paper II \citep{lester19b}, and Paper III \citep{lester20} -- in order to determine the fundamental stellar parameters of the components. Precise fundamental parameters of binary systems are essential for determining orbital demographics \citep{raghavan10, bordier22}, testing  models of stellar structure and evolution \citep[e.g.,][]{claret18, morales22}, calibrating mass and distance determination methods for single stars \citep{torres10, chaplin13, gallenne18}, and studying how binary stars form and evolve \citep[e.g.,][]{richardson21}. In this paper, we present the results for the more massive binaries, HD~61859, HD~89822, HD~109510, and HD~191692. Higher mass stars evolve rather quickly, so matching the observed parameters of both components at the same age becomes more challenging for evolutionary models. 

HD~61859 (HR 2962, HIP 37580) consists of a pair of F-type stars in a 32 day orbit. The first spectroscopic orbit was measured by \citet{harper26}, then more recently updated by \citet{tomkin08} using high resolution echelle spectra. 

HD~89822 (ET UMa, HR 4072, HIP 50933) is a well-studied binary consisting of an early A-type star and an early F-type star in a 12 day orbit. Spectroscopic orbits were determined by \citet{schlesinger12} and \citet{nariai70}. The primary component of HD~89822 was identified as a Mercury-Manganese (HgMn) star \citep{abt73}, the higher mass cousins of metallic line (Am) stars. \citet{adelman94} completed a detailed abundance analysis of the system to confirm this classification and found that the secondary is an Am star \citep[see also][]{chojnowski20}. The TESS light curve of HD~89822  also shows a heartbeat pattern \citep{kochukhov21, kolaczek21}, which is a brightening near periastron due to tidal distortion in close eccentric binaries \citep{thompson12}, but the system shows no evidence of the rotational modulation found in some HgMn stars \citep{kochukhov21}. %(amp ~ 0.3 mmag)

HD~109510 (24 Com B, HR 4791, HIP 61415) contains a pair of A-type stars in a 7 day orbit. HD~109510 has been studied spectroscopically by \citet{petrie37} and \citet{mayor87}. HD~109510 also has a bright visual companion, HD~109511 (24 Com A), at a separation of 20\arcsec. This star has a similar proper motion, radial velocity, and parallax to HD~109510, so they are likely physically associated. However, HD~109511 is beyond the fields-of-view of our telescopes and will not affect our observations. The timescales of the inner orbits are also much shorter than that of the outer pair, so HD~109511 would not cause visible perturbations in the orbit of HD~109510.

HD~191692 ($\theta$ Aql, HR 7710, HIP 99473) contains a pair of late B-type stars in a 17 day orbit. A spectroscopic orbit was measured by \citet{cesco46} and a visual orbit was measured by \citet{hummel95} using the Mark III interferometer; \citet{pourbaix00} later used these observations to determine a combined (VB+SB2) orbital fit. However, these observations lack the high spectral and angular resolution of modern day echelle spectrographs and long baseline interferometers, so new observations will provide more precise results. In addition, a detailed abundance analysis of HD~191692 was completed by \citet{adelman15}, which found the primary component to have solar abundances and the secondary component to be weakly metallic lined.  

In Section~\ref{section:spec}, we present our spectroscopic observations and radial velocities. In Section~\ref{section:inter}, we describe our interferometric observations, binary position measurements, and combined orbital solution. In Section~\ref{params}, we determine the fundamental stellar parameters of each component and compare the results to stellar evolution models. Finally, we discuss our results in Section~\ref{discussion}.

% -----------------------------------------------------------------------------
\section{Spectroscopy}\label{section:spec}

\subsection{APO Observations}
We observed HD~61859, 89822, and 109510 with the ARC echelle spectrograph \citep[ARCES;][]{arces} on the APO 3.5~m telescope from 2015--2020. ARCES covers 3500-10500 \AA\ over 107 orders at an average resolving power of $R\sim30,000$. We reduced our data using standard echelle procedures in IRAF, then removed the blaze function using the procedure in Appendix~A of \citet{kolbas15}. Radial velocities ($V_{r 1}$, $V_{r 2}$) were measured with the multi-order TODCOR method \citep{todcor1, todcor2} as described in Paper~II. In summary, TODCOR calculates the cross correlation function (CCF) for a grid of primary and secondary radial velocities. The CCFs for each echelle order were added together to find the maximum CCF, corresponding to the best-fit radial velocities and their uncertainties ($\sigma_1, \sigma_2$). Templates were created with the use of BLUERED model spectra \citep{bluered} and atmospheric parameters from recent literaure. Our observations are listed in Table~\ref{rvtable} in the Appendix with the UT date, HJD date, radial velocity and uncertainty of each component, and the telescope used. TODCOR also calculates the monochromatic flux ratio ($f_2/f_1$) near H$\alpha$, which we use later to estimate the radius ratio for each system (see Section~\ref{sedfit}). 

\subsection{CTIO Observations}
We observed HD~109510 and 191692 with the CHIRON echelle spectrograph \citep{chiron, paredes21} on the CTIO/SMARTS 1.5m telescope from 2014--2020. HD~109510 was observed in the $R\sim28,000$ fiber mode and HD~191692 was observed in the $R\sim90,000$ slicer mode. Both modes cover 4500-8800 \AA\ over 60 echelle orders. The data were reduced with the CHIRON team's pipeline, then we used the above procedure to remove the blaze function and measure radial velocities of each component. These observations are also listed in Table~\ref{rvtable}.

\subsection{Fairborn Observations} 
We also acquired spectroscopic observations of all four binaries at Fairborn Observatory in southeast Arizona with the Tennessee State University 2.0 m Automatic Spectroscopic Telescope (AST) and a fiber-fed echelle spectrograph \citep{eaton04}. We obtained spectra of HD~61859 from 2011--2020 as a continuation of the velocities published by \citet{tomkin08}. We collected spectra of HD~89822 from 2005--2020, HD~109510 from 2020--2021, and of HD~191692 from 2004--2021. The observations of HD~89822 and HD~191692 that were obtained before 2011 were acquired with a 2048 $\times$ 4096 SITe ST-002A CCD. Those spectra have 21 orders, cover a wavelength region of 4920--7100~\AA, and have $R\sim35,000$ at 6000~\AA. During the summer of 2011 we replaced the SITe CCD with a Fairchild 486 CCD that has a 4096 $\times$ 4096 pixel array enabling coverage of a wavelength range of 3800--8600~\AA\ over 48 orders \citep{fekel13}. We used a 200~$\mu$m fiber that produced $R\sim25,000$ at 6000~\AA. \citet{eaton07} explained the data reduction and wavelength calibration of the raw AST spectra.

\citet{fekel09} provided a general description of the typical velocity reduction. Briefly, for HD~61859 and HD~109510 we used a solar-type star line list consisting of 168 mostly neutral Fe lines in the wavelength region 4920--7100~\AA. The early-A and late-B spectral classes of the components of HD~89822 and HD~191692 required the use of our A-star line list. The lines in that list cover the same wavelength region as the solar-type line list but mainly include ionized Fe lines. Each line was fitted with a rotational broadening function \citep{sandberg11}, and if the lines of the two components were partly blended we obtained a simultaneous fit. The stellar velocity was determined as the average of the line fits. A value of 0.3 km~s$^{-1}$ was added to the SITe CCD velocities and 0.6 km~s$^{-1}$ to the Fairchild CCD velocities to make the resulting velocities from the two CCDs consistent with the velocity zero point of \citet{scarfe10}. These observations are also listed in Table~\ref{rvtable}.

\subsection{Preliminary Spectroscopic Orbit} \label{sborbit}
Following the procedure in previous papers, we accounted for differences in the zero-point offset of each spectrograph by first fitting separate orbital solutions to each data set using the RVFIT program\footnote{\href{http://www.cefca.es/people/~riglesias/rvfit.html}{http://www.cefca.es/people/$\sim$riglesias/rvfit.html}} \citep{rvfit}. We solved for the six spectroscopic orbital parameters of each system: the orbital period ($P$), epoch of periastron ($T$), eccentricity ($e$), longitude of periastron of the primary star ($\omega_1$), velocity semi-amplitudes ($K_1$, $K_2$), and systemic velocity ($\gamma$). We also fit an orbit to the previously published velocities of HD~61859 from \citet{tomkin08} using preliminary uncertainties equal to $1/\sqrt{weight}$. We shifted the APO, CTIO, and literature data sets such that the systemic velocities match those of the Fairborn data sets. We then used the $\chi^2$ values from the individual solutions to rescale the uncertainties such that the reduced $\chi^2$ of each data set equals one. Our adjusted velocities and rescaled uncertainties are those listed in Table \ref{rvtable} in the Appendix.

% -----------------------------------------------------------------------------
\section{Interferometry}\label{section:inter}

\input{table_relpos.txt}

\subsection{CHARA Array Observations} 
We observed all four binaries with the CHARA Array from 2012--2019 using the CLIMB \citep{climb} beam combiner and from 2019--2021 using the MIRC-X \citep{anugu20} beam combiner. CHARA has six 1.0~m telescopes arranged in a Y-shape with baselines ranging from 34--331~m \citep{chara}. CLIMB combines the $K'$-band light from three telescopes at a time, while MIRC-X combines the $H$-band light from up to six telescopes and disperses the light into six narrowband spectral channels at $R=50$. Table~\ref{relpos} lists the UT date, HJD, and beam combiner for each CHARA observation. The CLIMB data were reduced with the pipeline developed by John D. Monnier;  the general method is described in \citet{monnier11} and the extension to three beams is described in \citet{kluska18}.  The MIRC-X data were reduced using the pipeline (version 1.3.3--1.3.5) developed by Jean-Baptiste Le~Bouquin and the MIRC-X team\footnote{\href{https://gitlab.chara.gsu.edu/lebouquj/mircx_pipeline.git}{https://gitlab.chara.gsu.edu/lebouquj/mircx\_pipeline.git}}, which splits each 10~min data sequence into four 2.5~minute bins. These reductions produce squared visibilities ($V^2$) for each baseline and closure phases (CP) for each closed triangle. We corrected for any instrumental and atmospheric effects on the observed visibilities using observations of calibrators stars, whose uniform disk angular diameters were taken from SearchCal \citep{searchcal} and are listed in Table~\ref{cal} in the Appendix. On each night, we also calibrated the calibrators against each other and did not find evidence of binarity in the calibrators.

\input{table_orbpar.txt}

\subsection{Binary Positions}  
We measured the relative positions from the visibilities and closure phases using the method\footnote{\href{http://www.chara.gsu.edu/analysis-software/binary-grid-search}{http://www.chara.gsu.edu/analysis-software/binary-grid-search}} of \citet{schaefer16}, which searches across a grid of separations in right ascension and declination to find the best-fit relative position. At each grid point, we compared the observed $V^2$ and CP to model values to fit for the flux ratio and calculate the $\chi^2$ value. We then searched a small area around the best-fit position to find the contour marking $\chi^2 \le \chi^2_{min}+1$ that determines the major axis ($\sigma_{maj}$), minor axis ($\sigma_{min}$), and position angle ($\phi$) of each error ellipse. Because the orbital periods of these systems are much longer than the observation time, any small orbital motion within a single night is contained within the error ellipses. The best-fit relative positions, error ellipse parameters, and flux ratio estimates for each night are listed in Table~\ref{relpos}. 

We note that the flux ratios are likely underestimated, because they do not account for systematics in visibility miscalibrations. Furthermore, the more precise CP of MIRC-X can constrain the flux ratios better than CLIMB, so we used the flux ratios from the MIRC-X observations to estimate the radius ratios in Section~\ref{sedfit}. We calculated the uncertainty for a single measurement from the standard deviation of the flux ratios derived from the three MIRC-X observations of HD 191692. We used this value as the final uncertainty for HD 61859, HD 89822, and HD 109510, but scaled this value by $\sqrt3$ to calculate the final uncertainty for HD 191692.

For HD~191692, the primary star is large enough to be resolved by CHARA, so we simultaneously fit the $V^2$ and CP for the primary angular diameter ($\theta_1$) and the binary parameters. The secondary component is unresolved, so we held the angular diameter fixed to $\theta_{2} = 0$ mas. We found the weighted-mean, uniform-disk angular diameter to be $\theta_1 = 0.571\pm0.017$ mas from our MIRC-X observations. We also tested limb-darkened models using linear limb-darkening coefficients from \citet{claret00} for the effective temperature and surface gravity found in Section~\ref{sedfit}. However, the difference between the limb-darkened visibilities and the uniform-disk visibilities were roughly ten times smaller than the observational uncertainties, because limb-darkening is very weak in the near infrared \citep{davis00}. 

\subsection{Combined Visual + Spectroscopic Solution}\label{vbsbfit}
\begin{figure*}
\centering
\includegraphics[width=0.49\textwidth]{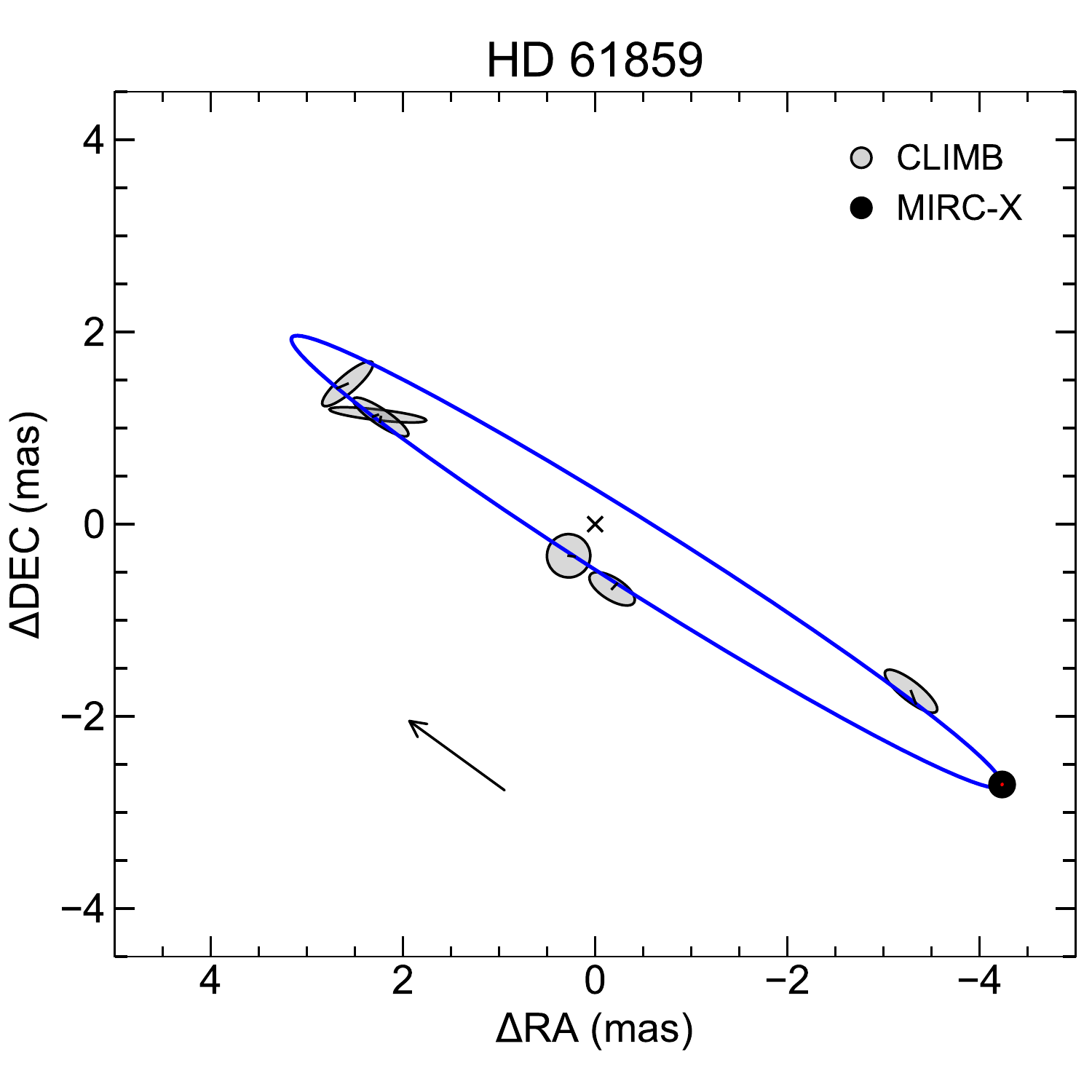} 
\includegraphics[width=0.49\textwidth]{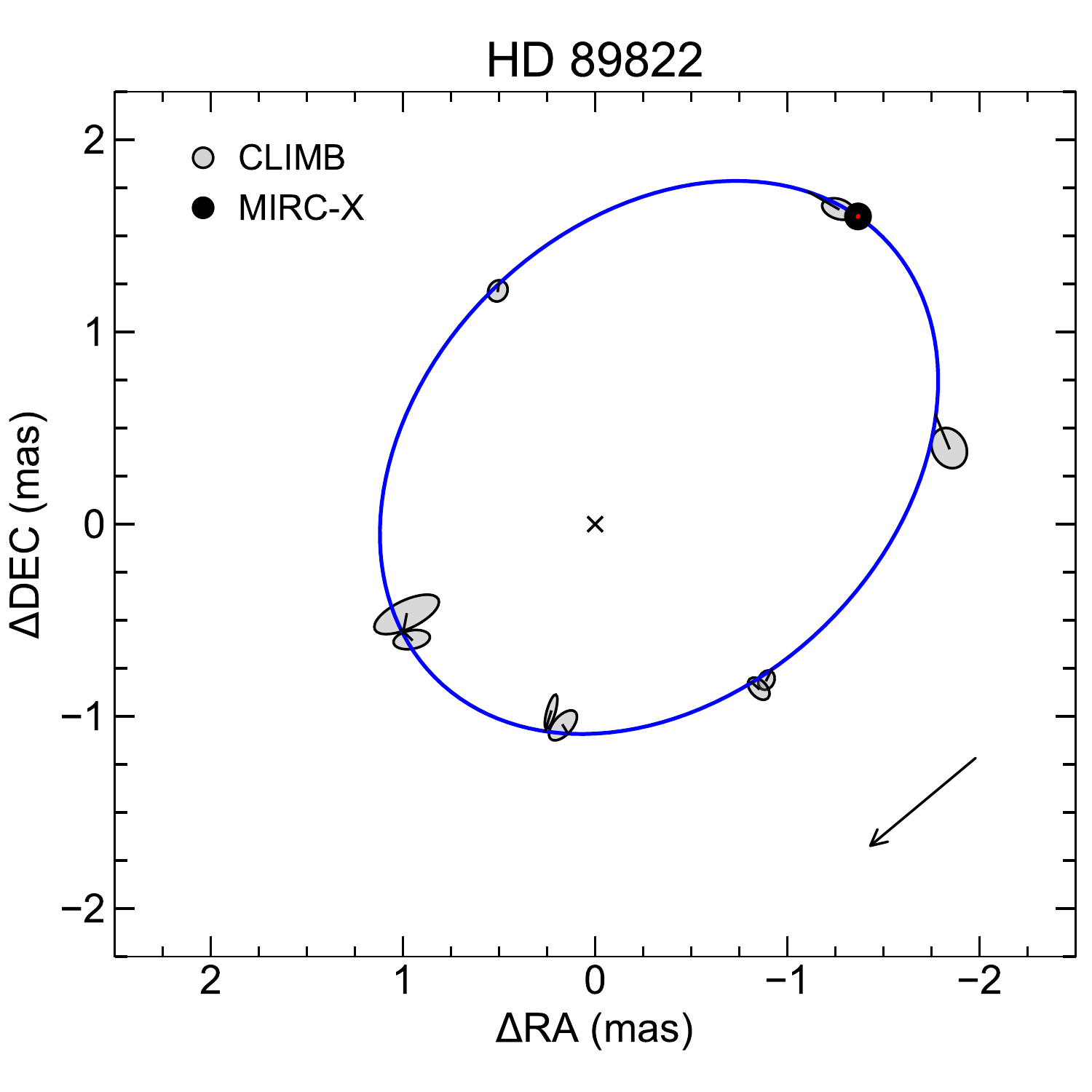} 
\includegraphics[width=0.49\textwidth]{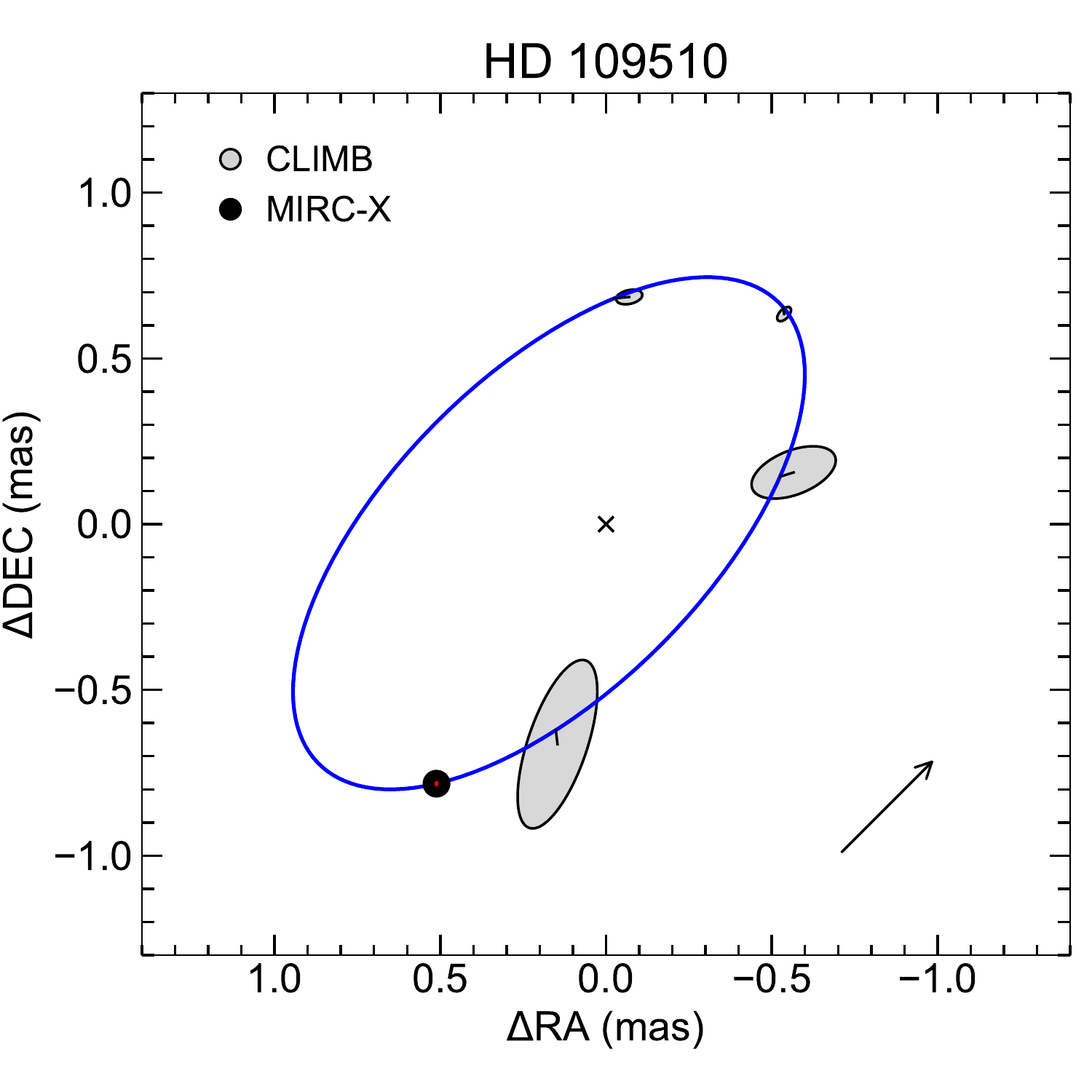} 
\includegraphics[width=0.49\textwidth]{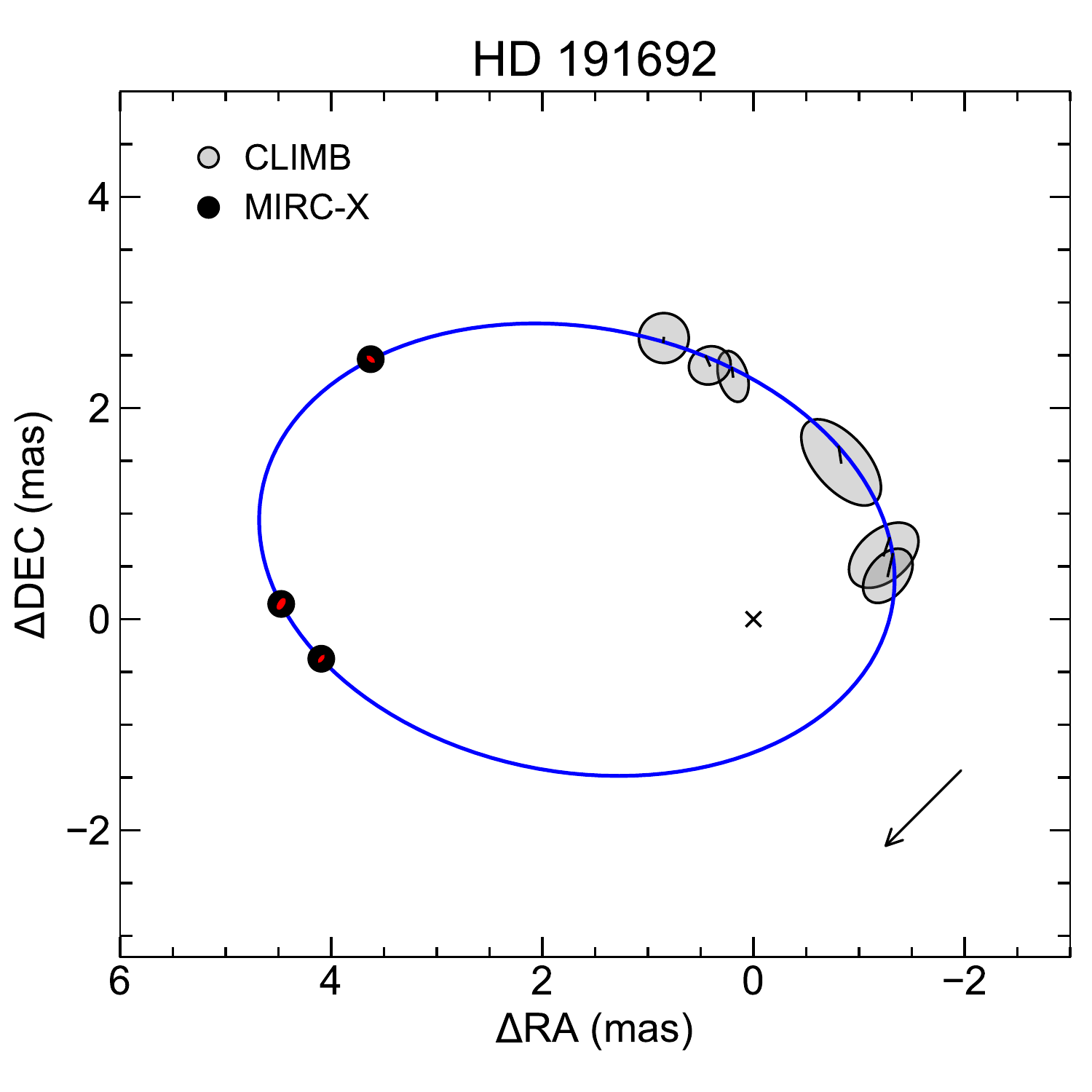} 
\caption{Visual orbits of HD~61859 (top left), HD~89822 (top right), HD~109510 (bottom left), and HD~191692 (bottom right). The primary star is located at the origin (black cross). The relative positions of the secondary star from CLIMB data are marked by the gray points corresponding to the sizes of the error ellipses. The relative positions from MIRC-X data are marked by black circles with red error ellipses, which are often to small to be seen. The solid blue curves represent the best-fit model visual orbits, and thin black lines connect each observed and model position. The arrows indicate the directions of orbital motion. \label{vborbit}}
\end{figure*}

\begin{figure*}
\centering
\includegraphics[width=0.49\textwidth]{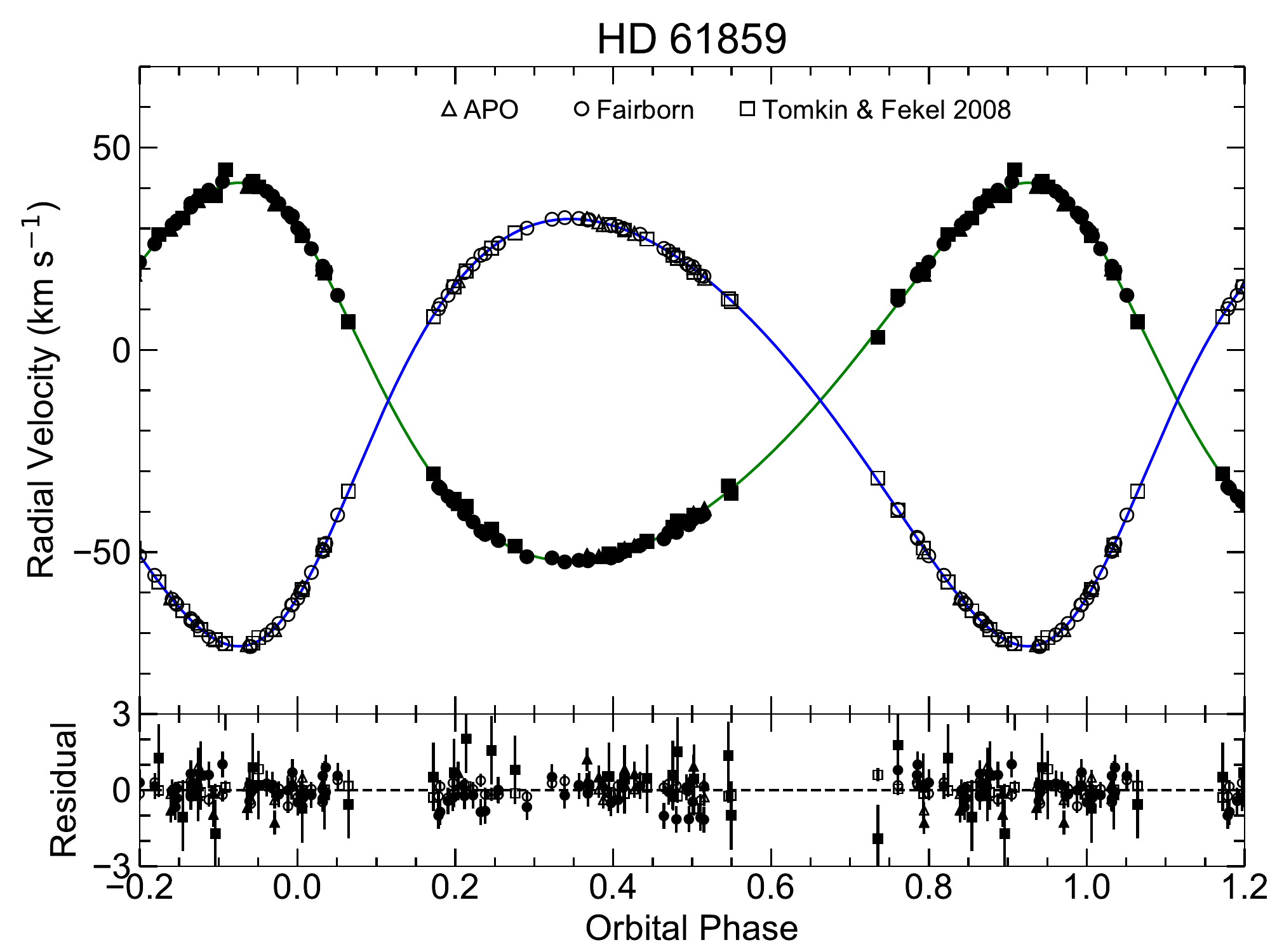} 
\includegraphics[width=0.49\textwidth]{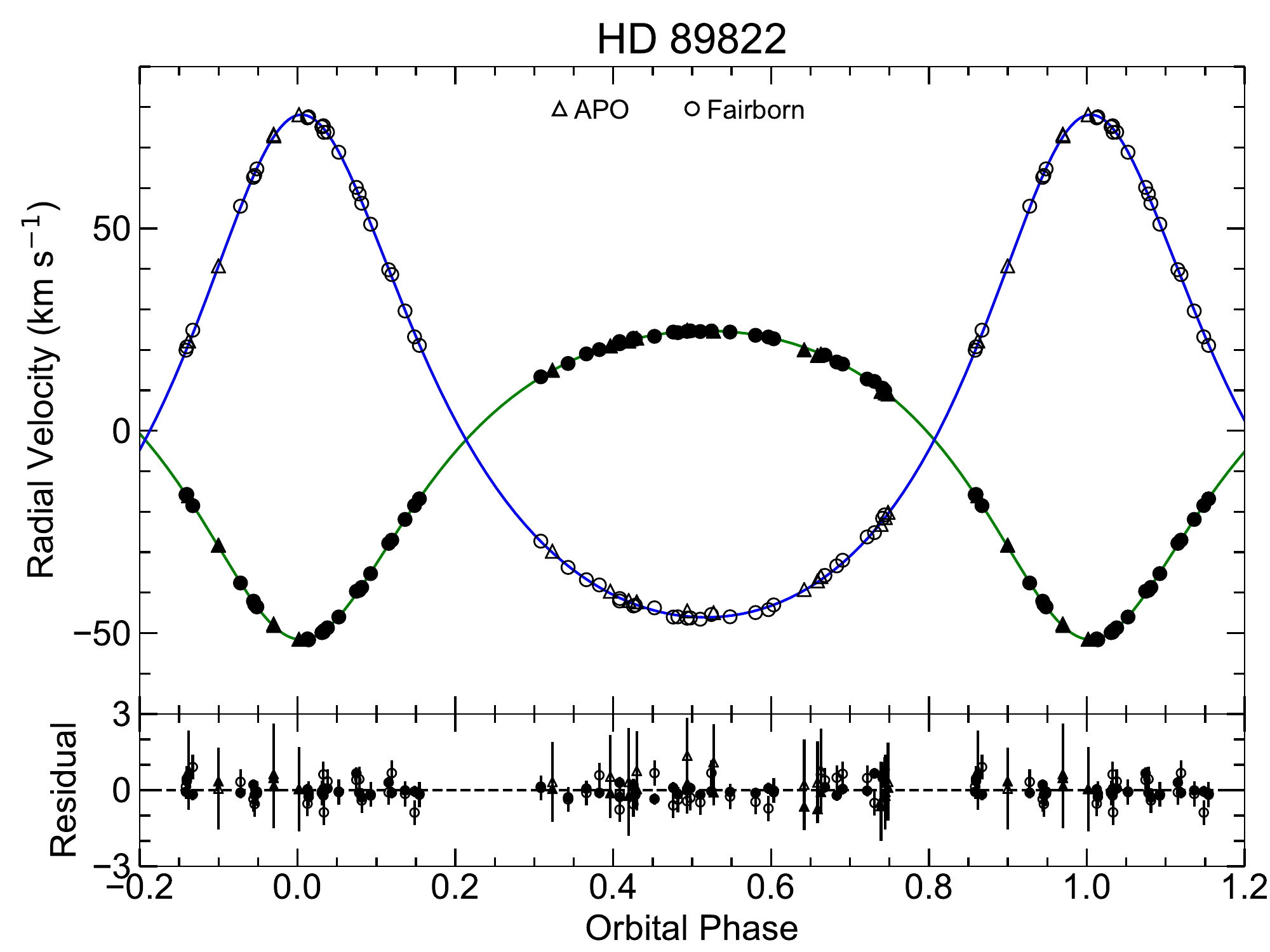} 
\includegraphics[width=0.49\textwidth]{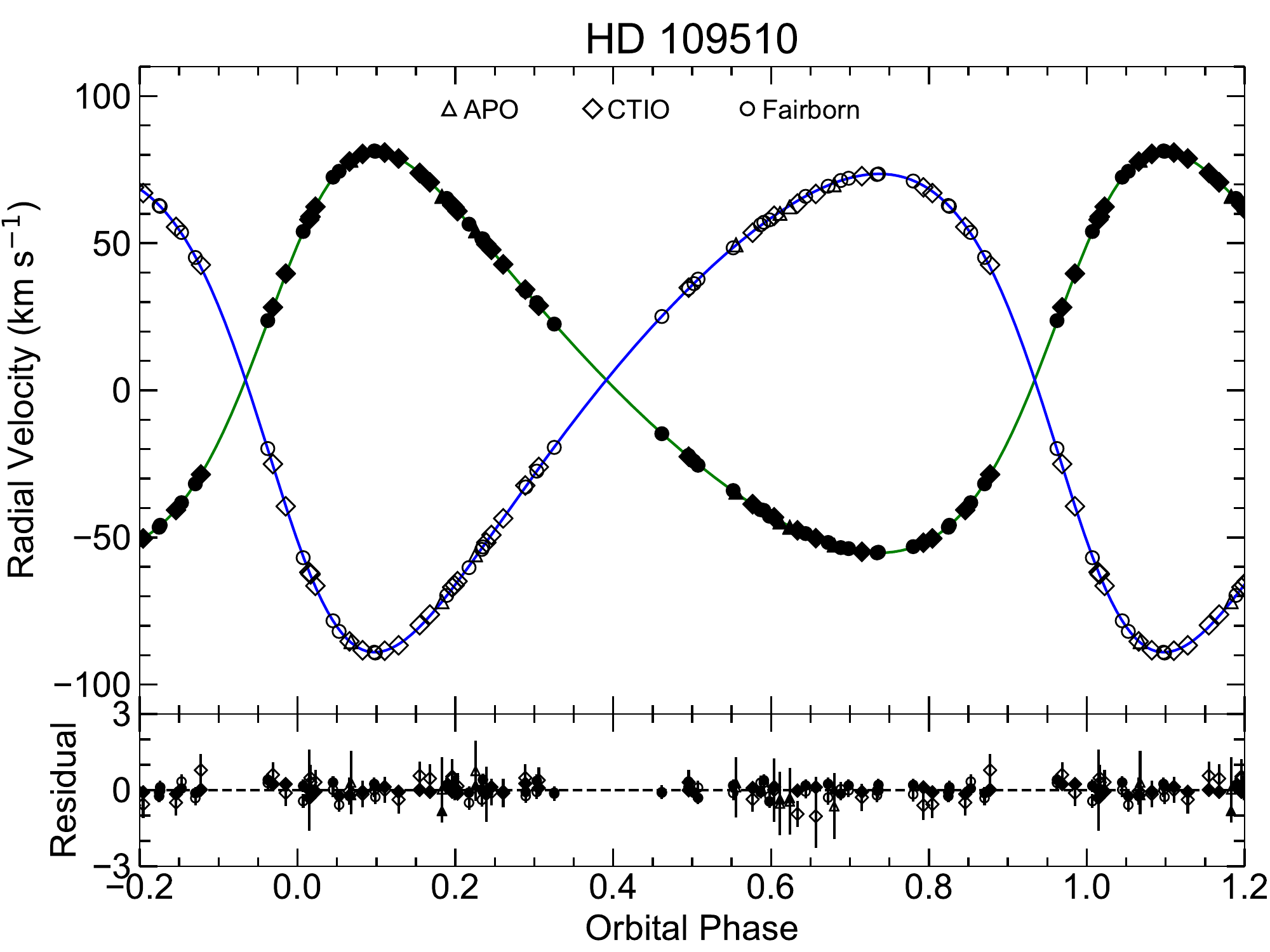} 
\includegraphics[width=0.49\textwidth]{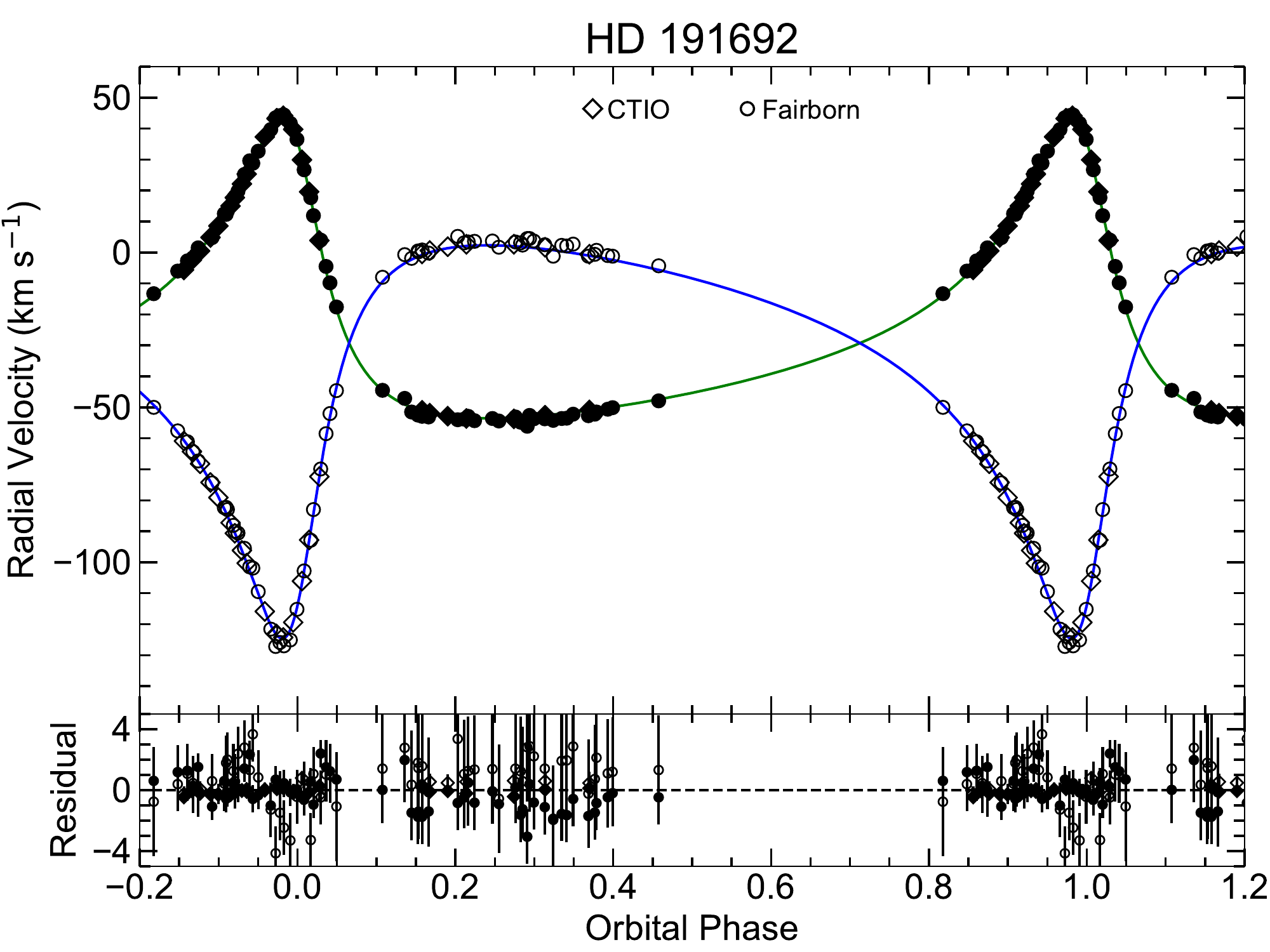} 
\caption{Radial velocity curves of HD~61859 (top left), HD~89822 (top right), HD~109510 (bottom left), and HD~191692 (bottom right). The observed data for the primary and secondary star are shown with the filled and open points, respectively. The triangles, diamonds, and circles represent the APO, CTIO, and Fairborn data, respectively. The model curves are shown with the solid lines, and the residuals to the fit are shown in the bottom panels.  \label{rvcurve}}
\end{figure*}

We determined the final orbital solution by simultaneously fitting the interferometric and spectroscopic data using the method\footnote{\href{https://www.chara.gsu.edu/analysis-software/orbfit-lib}{https://www.chara.gsu.edu/analysis-software/orbfit-lib}} of \citet{schaefer16}. The full set of orbital parameters includes the orbital period ($P$), epoch of periastron ($T$), eccentricity ($e$), longitude of periastron of the primary star ($\omega_1$), the inclination ($i$), the angular semi-major axis ($a$), the longitude of the ascending node ($\Omega$), the systemic velocity ($\gamma$), and the velocity semi-amplitudes ($K_1$, $K_2$). Table~\ref{orbpar} lists the combined (VB+SB2) orbital solutions for each system. The visual orbits of all four systems are shown in Figure~\ref{vborbit} and the spectroscopic orbits are shown in Figure~\ref{rvcurve}. To determine the uncertainty of each orbital parameter, we performed a Monte Carlo error analysis in which we varied each data point within its Gaussian uncertainty and refit for the orbital solution. We then made a histogram of the best-fit parameters from $10^5$ iterations and fit Gaussians to each distribution to determine the $1\sigma$ uncertainties in each parameter (also listed in Table~\ref{orbpar}).

% -----------------------------------------------------------------------------
\section{Stellar Parameters}\label{params}

\subsection{Masses and Distance}
We calculated the component masses and binary distance from the combined orbital solution using nominal Solar values from \citet{prsa16}. Our results are listed in Table~\ref{atmospar}. 
The uncertainties in component mass are less than 
1.4\% for HD~61859, 
5.5\% for HD~89822, 
11.9\% for HD~109510, and 
1.4\% for HD~191692. 
The uncertainties in distance are 
0.5\% for HD~61859, 
2.0\% for HD~89822, 
5.6\% for HD~109510, and 
0.4\% for HD~191692.

Model independent distance measurements from binary stars are important checks for other distances measurements, such as the Cepheid period-luminosity relationship \citep{gallenne18} and trigonometric parallax \citep{halbwachs16, stassun18}. We compared our distances from orbital parallax to the distances from Gaia DR3 parallaxes \citep{gaia1, gaiaDR3}, which are listed at the end of Table~\ref{atmospar}. Our distances were consistent with Gaia's to within the $1\sigma$ uncertainties for HD~61859, HD~89822, and HD~109510. However, our distance for HD~191692 of $76.3\pm0.3$~pc does not match the Gaia distance of $70.1\pm2.3$~pc. HD~191692 is a very bright star ($G=3.2$ mag), so the trigonometric parallax could have been affected by saturation. 

% DR3 values-- 
% star, parallax, error, distance, error
% 61859   15.4317  0.04518    64.8016     0.1897
% 89822    9.6318  0.09530    103.8227    1.0272
% 109510   8.9477  0.03047    111.761     0.3806
% 191692  14.2687  0.46226    70.082      2.27     

\break

\input{table_atmospar.txt}

\subsection{Atmospheric Parameters}\label{tempfit}
We reconstructed the spectra of each binary component using the method of \citet{tom} as described in Papers I-III. Initial model templates were created using atmospheric parameters from the recent literature. We then fit for the effective temperature (\teff), projected rotational velocity ($V\sin i$), and metal abundance ($\log Z/Z_\odot$) of each component by cross-correlating a grid of BLUERED model spectra against the reconstructed spectra. Only echelle orders in the range $4500-6600$\AA\ that contain strong absorption lines were used. We determined the best-fit parameters of each component and their uncertainties from the CCF peak, as listed in Table~\ref{atmospar}. Example plots of the final reconstructed spectra and best-fit model spectra are shown in Figures~\ref{rec61859}-\ref{rec191692} in the Appendix. For HD 61859, we found both components to be slightly metal poor. We found HD~89822 to have a slightly metal rich primary and a roughly Solar abundance secondary, consistent with the iron abundances found by \citet{adelman94}. We also confirm that the primary component is a HgMn star and the secondary component is an Am star based on the reconstructed spectra, as shown in Figure~\ref{rec89822}. Both components of HD~109510 are also slightly metal poor, but the secondary is consistent with Solar metallicity within the uncertainties. Finally, we found both components of HD~191692 to be slightly metal rich, consistent with the iron abundances found by \citet{adelman15}.

\subsection{Radii and Surface Gravities}\label{sedfit}	

\begin{figure*}
\centering
\includegraphics[width=0.49\textwidth]{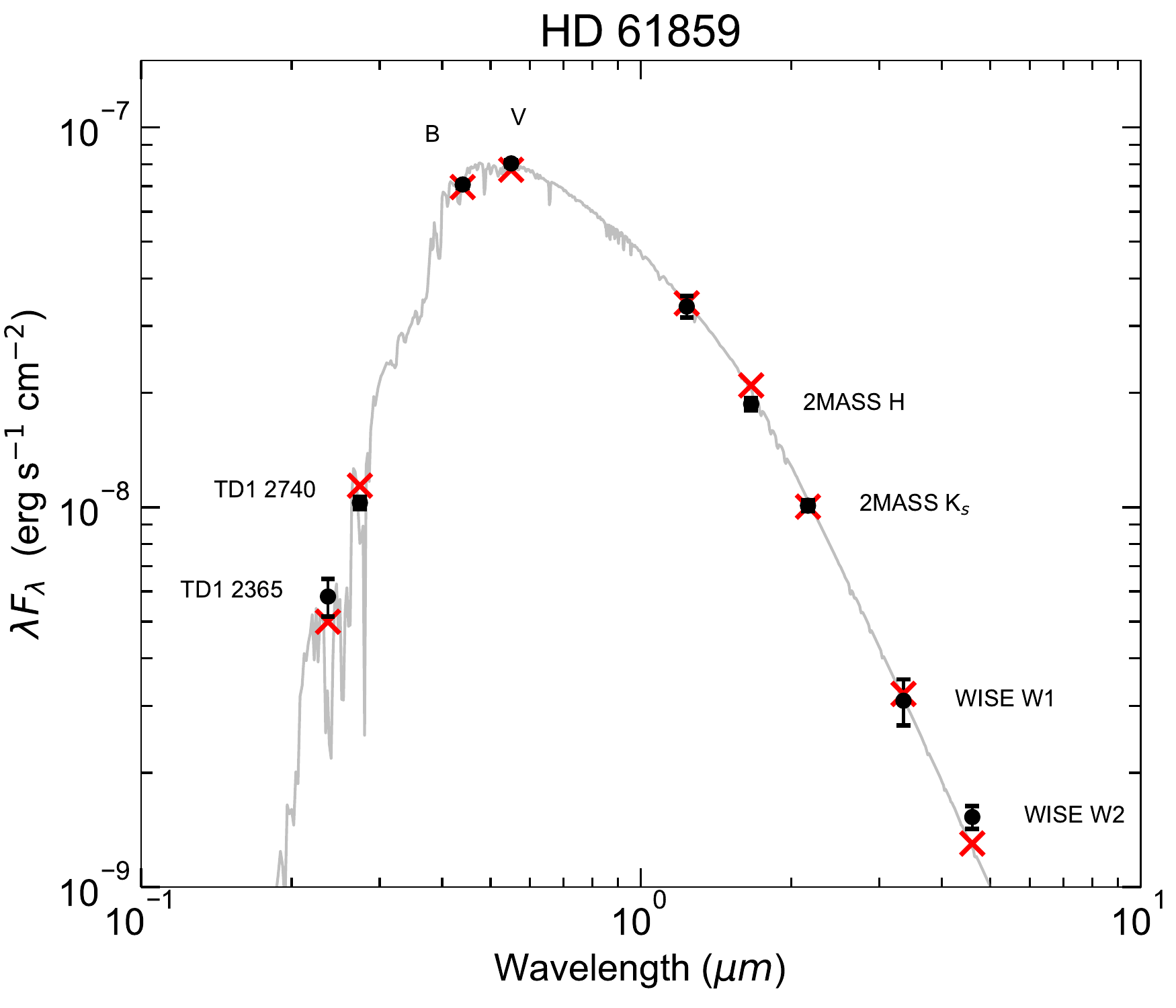}
\includegraphics[width=0.49\textwidth]{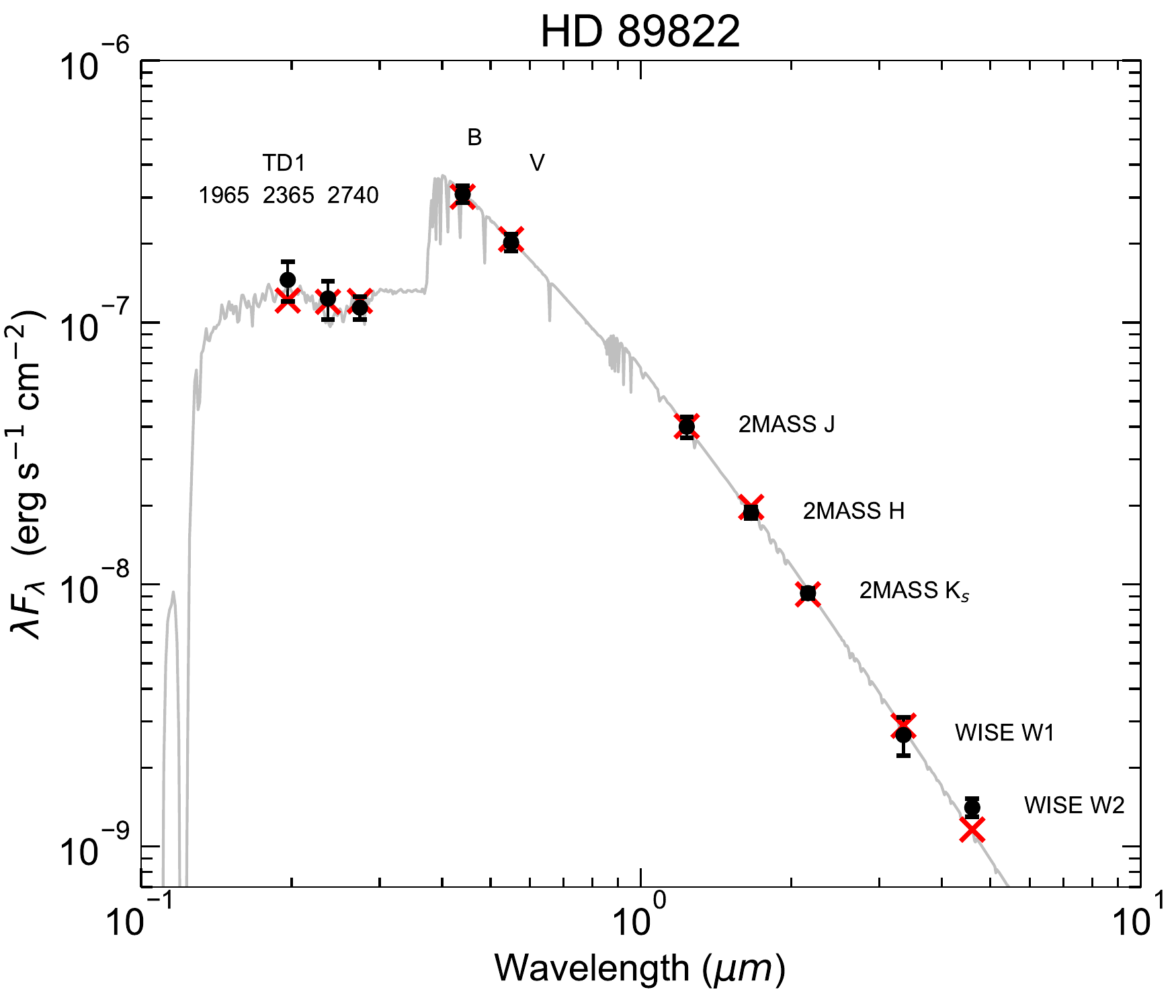}
\includegraphics[width=0.49\textwidth]{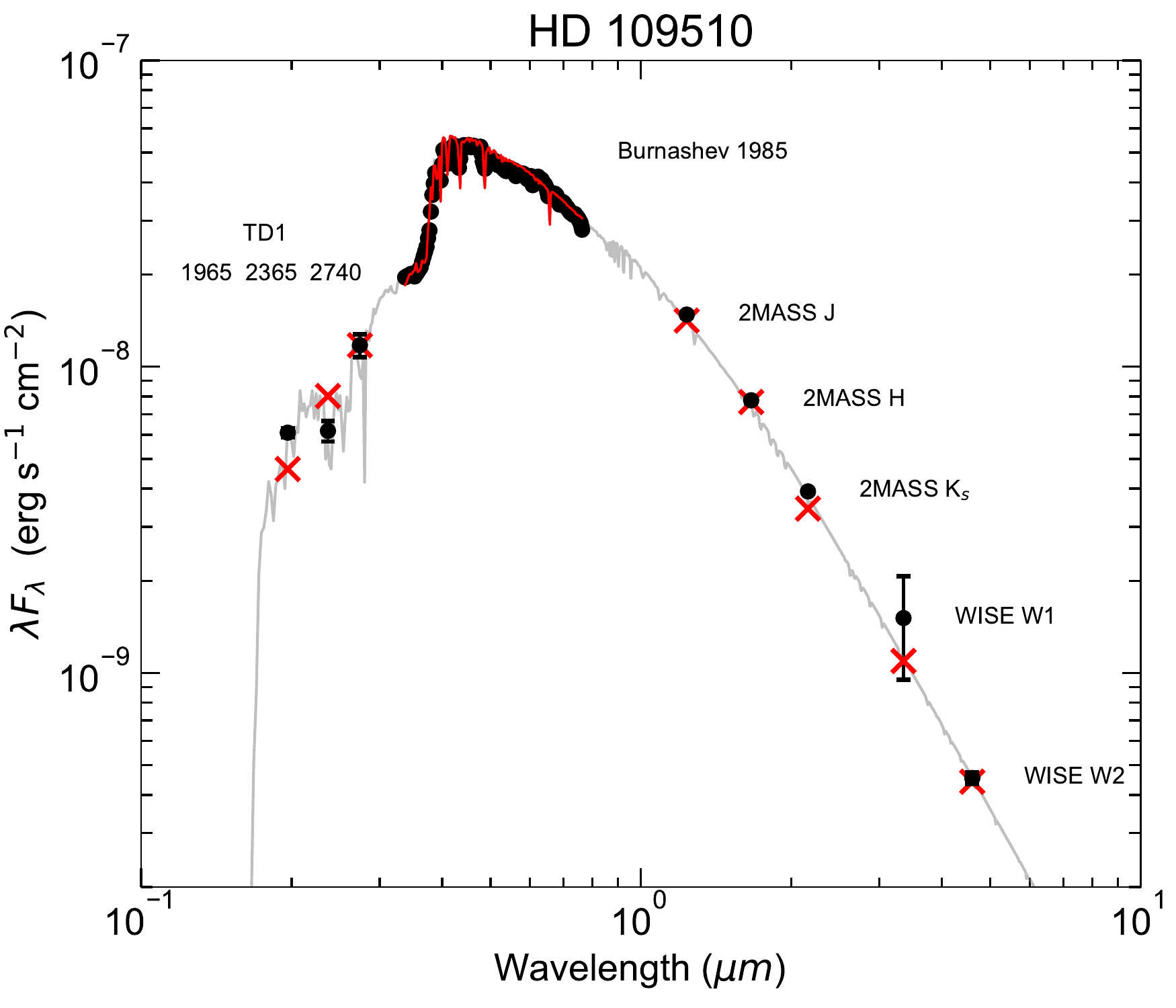} 
\includegraphics[width=0.49\textwidth]{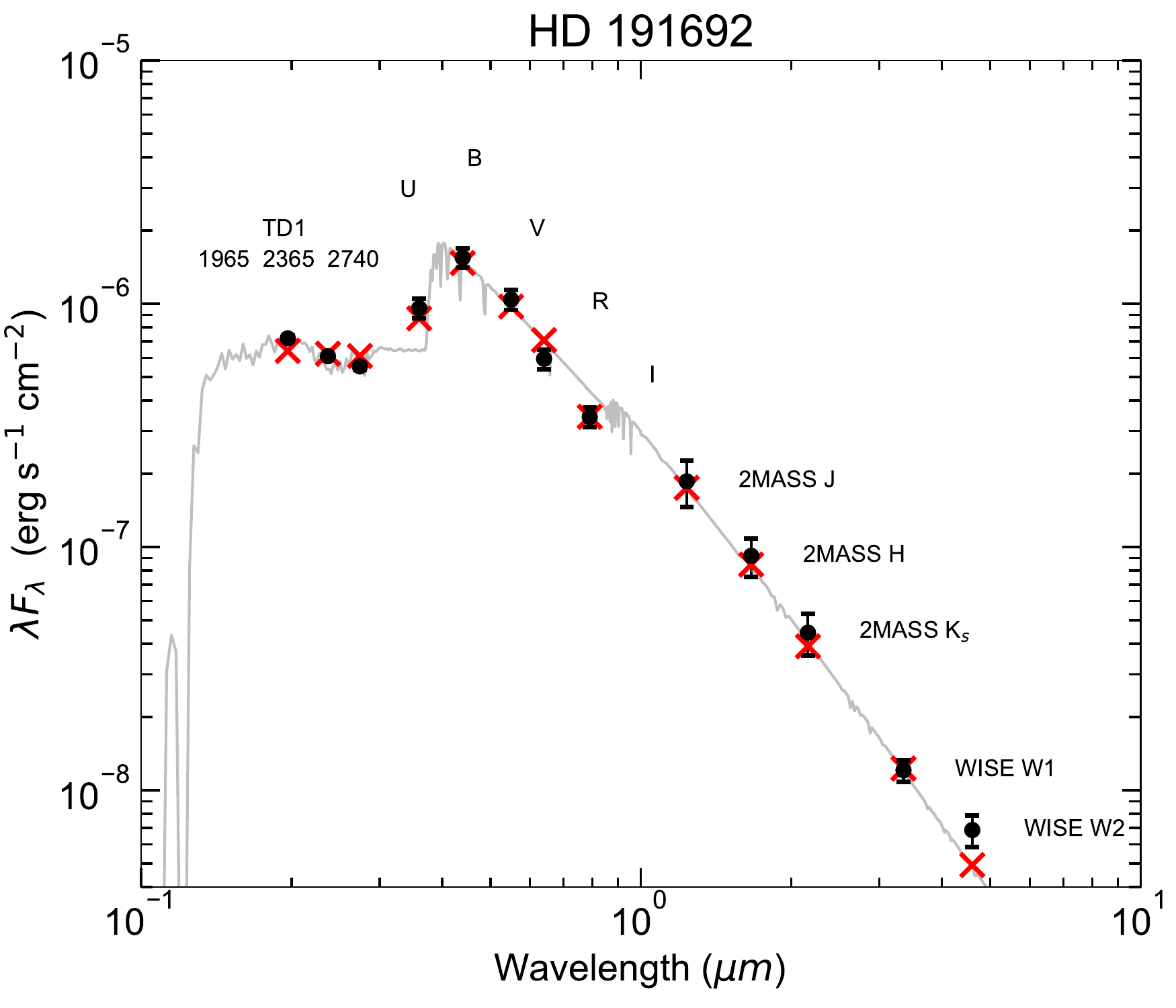} 
\caption{SEDs of HD~61859 (top left), HD~89822 (top right), HD~109510 (bottom left), and HD~191692 (bottom right). The observed fluxes are shown in black, and the best-fit binary model fluxes are shown as the red crosses. The full binary model is shown in gray. For clarity, we do not show the uncertainties in the \citet{burnashev85} spectra of HD 109510. \label{sed}}
\vspace{12pt}
\end{figure*}

We built spectral energy distributions (SEDs) using photometry from the literature in order to determine the radii of each component. We included UV photometry from the TD1 Stellar Ultraviolet Fluxes Catalog \citep{TD1}, optical photometry from the Fourth US Naval Observatory CCD Astrograph Catalog \citep[UCAC4;][]{ucac4}, and infrared photometry from the Two Micron All Sky Survey \citep[2MASS;][]{2mass} and the Wide-field Infrared Survey Explorer \citep[WISE;][]{wise}. The dataset for HD~109510 also includes low resolution, flux calibrated spectroscopy from \citet{burnashev85}.

Next, we created surface flux models of each component \citep{sedmodel} based on the \teff\ found above. We estimated the radius ratio ($R_2/R_1$) of each binary by comparing the model surface flux ratio and the observed flux ratios near $H\alpha$ (from spectroscopy) and in the $H$-band (from interferometry). These parameters are listed in Table~\ref{sedpar}. The surface flux models were also used to create a combined, binary SED model for each system.  Using the procedure described in Paper III, we fit the binary model SED to the observed fluxes in order to determine the angular diameter of the primary star ($\theta_1$) and the color excess ($E(B-V)$) of each system. For HD~191692, we converted the uniform-disk angular diameter of the primary component measured directly with interferometry to a limb-darkened disk diameter \citep{davis00}, then held this parameter fixed to fit only for reddening. We then used the radius ratios to calculate the angular diameters of the secondary stars ($\theta_2$). The observed SED's and best-fit models are shown in Figure~\ref{sed}, and the best-fit SED parameters are listed in Table~\ref{sedpar}. 

Finally, we combined the Gaia DR3 distances with the angular diameters to calculate the component radii, except in the case of HD~191692. The Gaia parallax of HD~191692 does not agree with our results and has a larger uncertainty, so we used the distance from orbital parallax instead. Table~\ref{atmospar} lists the best-fit stellar radii of each component, the surface gravities calculated from the masses and radii, and the luminosities calculated from the Stefan-Boltzmann Law. Our stellar radii have uncertainties less than 
4.0\% for HD~61859,
3.6\% for HD~89822, 
4.1\% for HD~109510, and 
3.0\% for HD~191692.

\input{table_sed.txt}

% -----------------------------------------------------------------------------
\subsection{Comparison with Evolutionary Models}\label{evofit}
We created evolutionary tracks for the components of each binary system using the Yonsei-Yale \citep{y2} and MESA \citep{mesa1, mesa5} codes in order to estimate the system ages and test the predictions of these models. We started with the observed masses and $\log Z/Z_\odot$ values from Table~\ref{atmospar}. The Yonsei-Yale models use a fixed overshooting parameter calculated from a step function in stellar mass. For the MESA models, we used the overshooting parameter ($f_{\rm ov}$) from the relation found by \citet{claret18} and the default mixing length parameter of $\alpha_{\rm ov}=2.0$, but adjusted $f_{\rm ov}$ and $\alpha_{\rm ov}$ slightly as needed to fit the observed parameters. Both models are non-rotating and use scaled Solar abundances. 

\begin{figure*}
\centering
\includegraphics[width=0.49\textwidth]{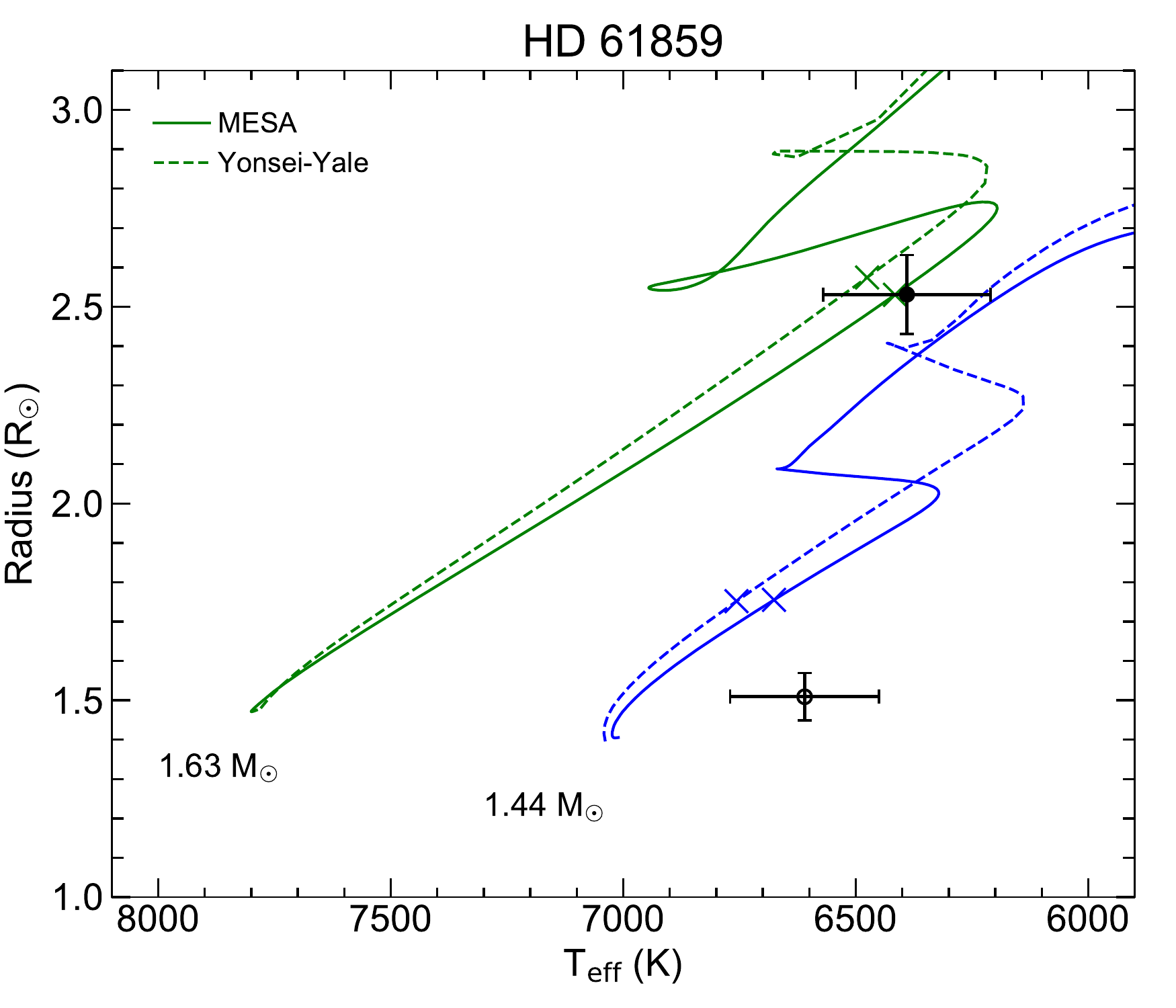}
\includegraphics[width=0.49\textwidth]{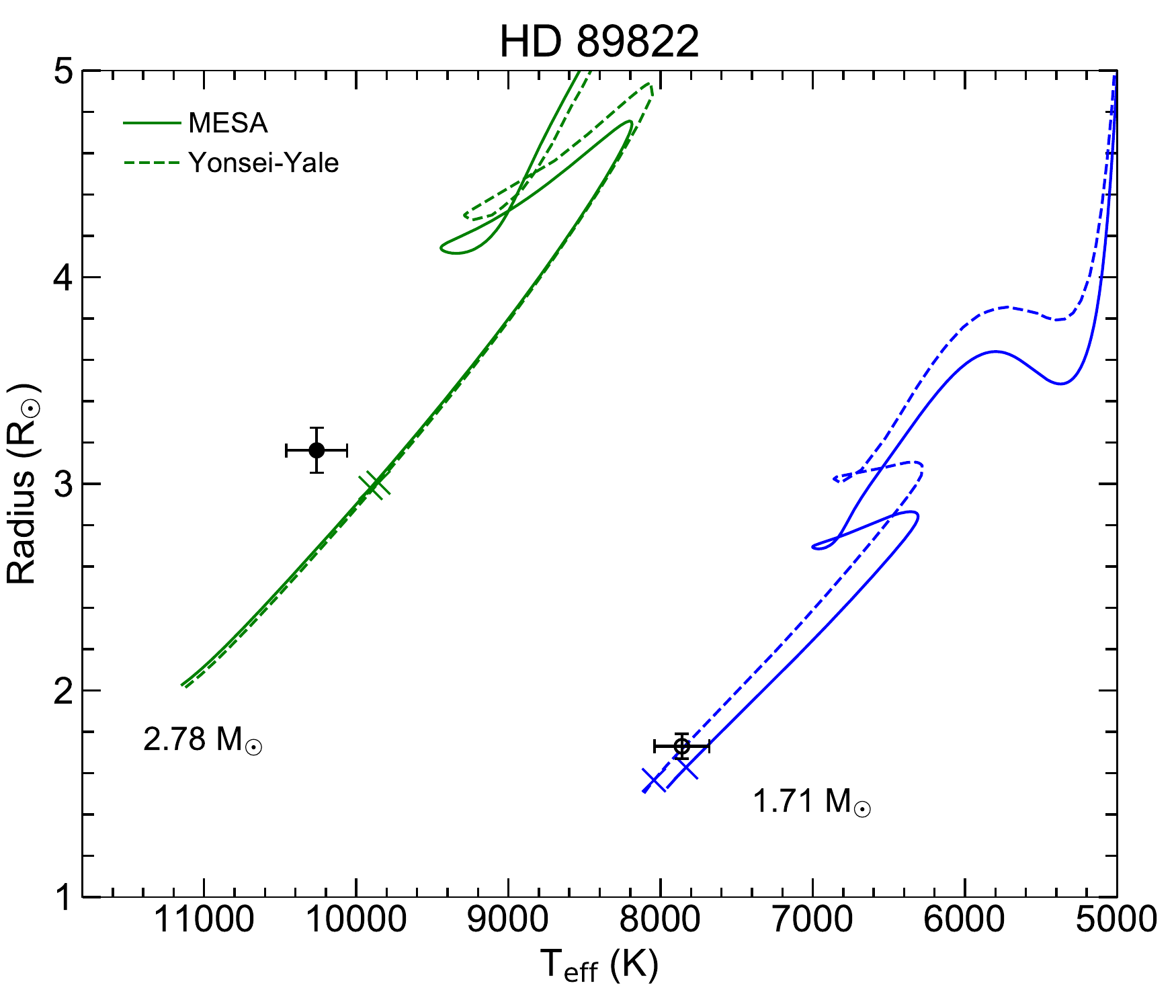}
\includegraphics[width=0.49\textwidth]{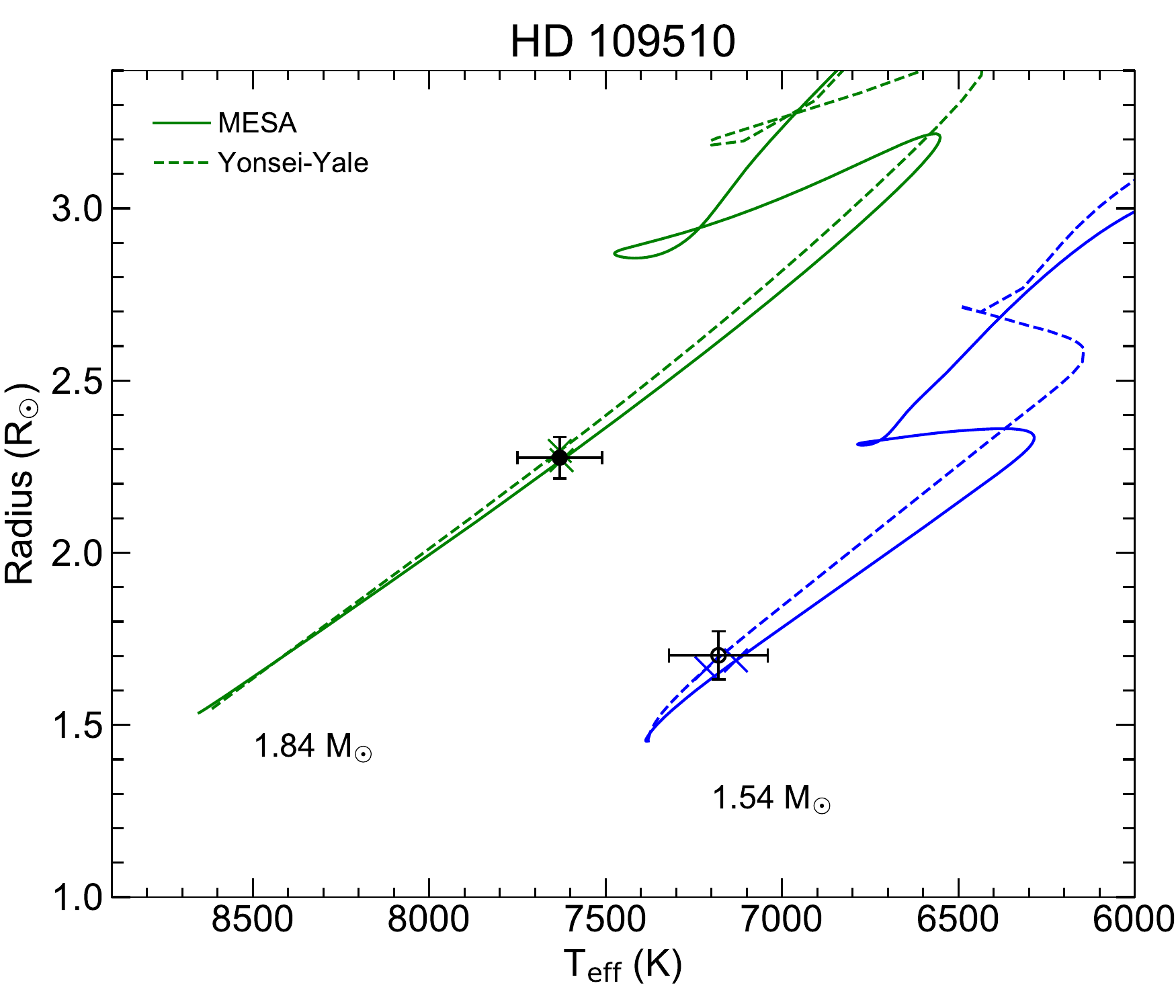}
\includegraphics[width=0.49\textwidth]{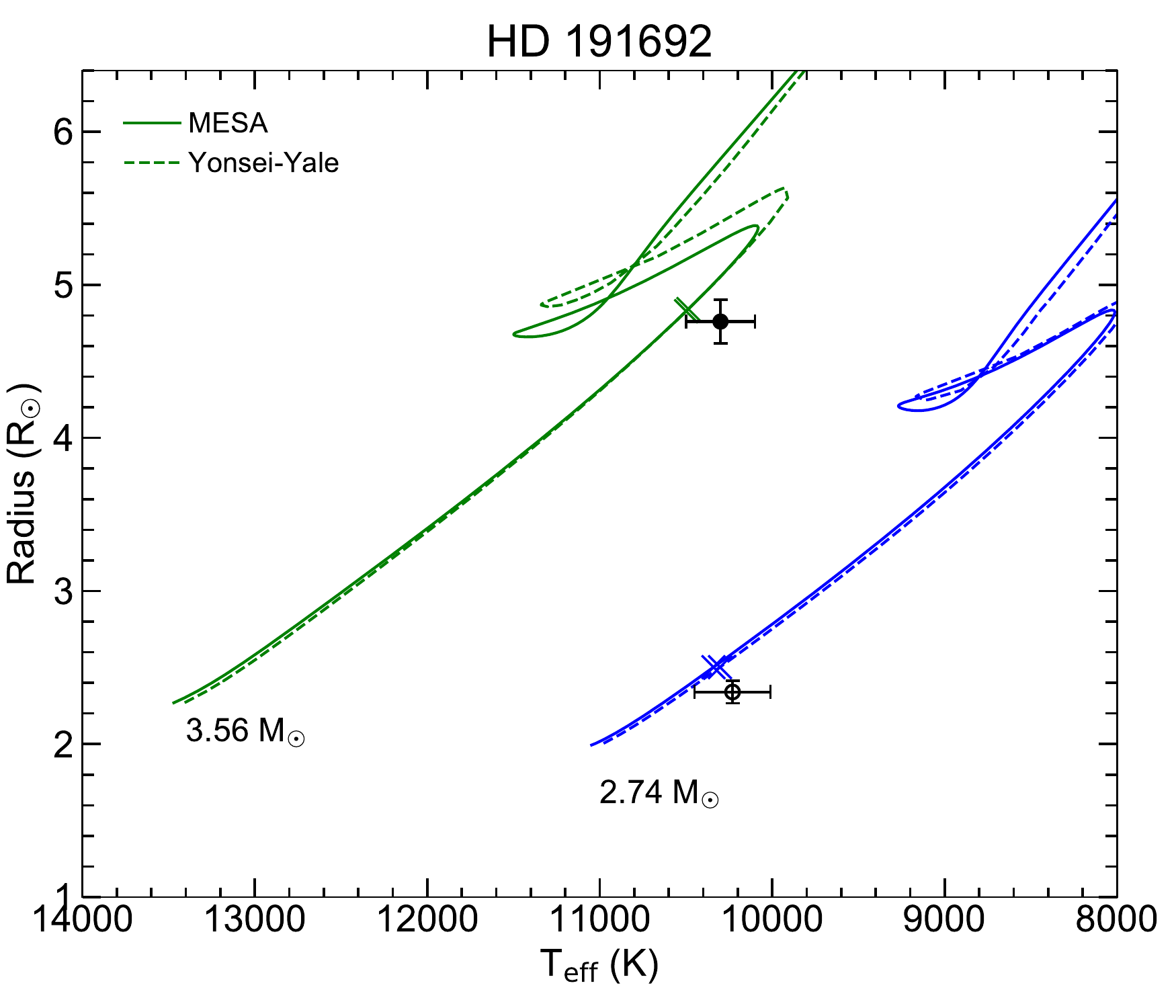}
\caption{Evolutionary tracks for HD~61859 (top left), HD~89822 (top right), HD~109510 (bottom left), and HD~191692 (bottom right). The observed stellar parameters are shown as the filled points for the primary stars and open points for the secondary stars. The Yonsei-Yale models are shown as dashed lines and the MESA models are shown as solid lines, with green for the primary and blue for the secondary. The crosses represent the position of the mean system age on each track. 
\label{evo}}
\vspace{12pt}
\end{figure*}

Figure~\ref{evo} shows the model evolutionary tracks and observed parameters of each binary system. For each component, we determined the age range where the model lies within the observed $1\sigma$ uncertainties. If the model lies outside the observed data point, we tried the $2\sigma$ uncertainties instead. If the component ages matched (to within 5\%), we took a weighted average to estimate the binary system age, where the mean values corresponded to the best-fit age of each component and the uncertainties corresponded to the spread in possible ages. As described below, the component ages often did not match, so we adopted the primary star's age as the system age because it evolves faster and provides a tighter age constraint. The estimated binary ages are marked on each evolutionary track by a cross in Figure~\ref{evo}.

\textit{HD~61859} --
The Yonsei-Yale models match the observed parameters of the primary component at age of 1.90 Gyr near the end of the main sequence, but lie just outside the observed parameters of the secondary at an age of 1.50 Gyr. The MESA models match the observed parameters of the primary component at age of 1.60 Gyr, but lie just outside the observed parameters of the secondary at an age of 1.10 Gyr. The observed parameters of the primary are not consistent with either model at the age of the secondary, so we adopted the primary star's mean age of 1.75 Gyr as the age for the system. A Solar-abundance evolutionary model would fit the secondary component better, despite being inconsistent with our spectroscopic analysis. Therefore, HD 61859 would benefit from a detailed abundance analysis to investigate whether or not the secondary component is chemically peculiar and solve this discrepancy.

\textit{HD~89822} --
The primary component is hotter and larger than predicted by both the Yonsei-Yale and MESA models. The closest position on the Yonsei-Yale tracks corresponds to 330 Myr, whereas the secondary component has an age of 510 Myr.  The closest position on the MESA tracks corresponds to 300 Myr, whereas the secondary component has an age of 400 Myr. We adopted the primary star's mean age of 315 Myr for the system, because the primary star would be too evolved at the predicted ages of the secondary star. The primary component likely does not match the models because it is a HgMn star; its abundance anomalies could bias our temperature determination from Section~\ref{tempfit} and our comparison to evolutionary models, because the BLUERED and MESA models both use scaled solar abundances. Because this system is quite young, we checked for membership in several nearby moving groups and associations using the \texttt{BANYAN $\Sigma$} tool \citep{gagne18}, but HD 89822 had a 99.9\% chance of being a field star.

\textit{HD~109510} --
The Yonsei-Yale models successfully fit both components at an age of 1.04 Gyr, and the MESA models successfully fit both components at an age of 930 Myr using $f_{\rm ov}$ from \citet{claret18} and the default $\alpha_{\rm ov}$ parameter. Even though this system has large uncertainties in stellar mass, our results are likely still accurate because the component effective temperatures and radii fit quite well.

\textit{HD~191692} -- 
Both components of HD~191692 are slightly cooler and smaller than predicted by the Yonsei-Yale and MESA models. The primary component has evolved to the end of the main sequence; the closest model track point corresponds to an age of 220 Myr for the Yonsei-Yale models and 210 Myr for the MESA models. The closest track points to the secondary correspond to an age of 170 Myr for the Yonsei-Yale models and 190 Myr for the MESA models. Even though these are very close to the ages of the primary star, the primary evolves much faster and would only be halfway up the main sequence at these ages. Therefore, we adopt the primary star's mean age of 215 Myr as the estimated age of the binary. Note that \citet{adelman15} derived similar component effective temperatures from their abundance analysis, so perhaps the observed radii are underestimated and causing this age discrepancy. While the primary star is just barely resolved by our MIRC-X observations, its angular diameter would be well above the resolution limit at shorter wavelengths, so future  interferometric observations would help solve this discrepancy between the observed and model parameters. Finally, we also checked if HD 191692 could be in young association using \texttt{BANYAN $\Sigma$}, but HD 191692 had a 99.9\% chance of being a field star.

% -----------------------------------------------------------------------------
\section{Discussion}\label{discussion}
We measured the visual and spectroscopic orbits for four binary systems using long baseline interferometry and high resolution spectroscopy. We constrained the stellar masses to $1-12$\% uncertainty, distances to $0.4-6$\% uncertainty, and stellar radii to $3-5$\% uncertainty. These fundamental parameters of longer period binaries ($P>7$d) are especially useful for testing our stellar models, because the component stars are less affected by tidal interactions and distortions and are better proxies for the evolution of single stars \citep{serenelli21}.

Typically, uncertainties in mass and radius less than 3\% are needed to test our models of stellar structure and evolution \citep{torres10}. HD~61859 and HD~191692 meet this criterion for stellar mass, while additional observations with long baseline interferometry would be needed to more precisely measure the orbital inclinations and stellar masses of HD~89822 and HD~109510. HD~191692 also meets this criterion for stellar radius, because the primary's angular diameter was measured more precisely with the CHARA Array. We estimate that both components of HD~61859, the primary component of HD~89822, and both components of HD~191692 are resolvable with CHARA in visible light, so future work to measure directly the component radii would greatly improve the comparison with stellar evolution models. 

We also found that HD~61859 is highly inclined and could create grazing eclipses. The TESS light curve potentially shows a very weak eclipse (0.1\% depth) when the secondary star passes front of the primary star. However, no complementary eclipses were seen, and each TESS sector only covers two-thirds of the binary orbit, so additional observations would be needed to confirm this system as an eclipsing binary.

Finally, the MESA models did not adequately fit the observed parameters of the chemically peculiar stars, HD~89822 and HD~191962, likely because their abundances are not scaled Solar values. \citet{abt61} found that the metallic line phenomenon is linked to binarity, so further study of these systems could improve our understanding of the structure and evolution of HgMn and Am stars.

% -----------------------------------------------------------------------------
\acknowledgments
The authors would like to thank the staff at APO, CHARA, CTIO, and Fairborn Observatory for their invaluable support during observations. 
K.V.L. is supported by an appointment to the NASA Postdoctoral Program at the NASA Ames Research Center, administered by Oak Ridge Associated Universities under contract with NASA.
%We would also like to thank the anonymous referee for their insightful comments. 
This work is based upon observations obtained with 
the Apache Point Observatory 3.5-meter telescope, owned and operated by the Astrophysical Research Consortium;
the Georgia State University Center for High Angular Resolution Astronomy Array at Mount Wilson Observatory, supported by the National Science Foundation under Grants No. AST-1636624, AST-1908026, \& AST-2034336; 
and the CTIO/SMARTS 1.5m telescope, operated
as part of the SMARTS Consortium.
Institutional support has been provided from the GSU College of Arts and Sciences and the GSU Office of the Vice President for Research and Economic Development. 
MIRC-X received funding from the European Research Council (ERC) under the European Union's Horizon 2020 research and innovation programme (Grant No. 639889). J.D.M. acknowledges funding for the development of MIRC-X (NASA-XRP NNX16AD43G, NSF-AST 1909165). S.K. acknowledges support from an ERC Consolidator Grant (Grant Agreement ID 101003096) and STFC Consolidated Grant (ST/V000721/1). 
Astronomy at Tennessee State University is supported by the state of Tennessee through its Centers of Excellence Program. 
This work has also made use of the Jean-Marie Mariotti Center \texttt{Aspro} \& \texttt{SearchCal} services, the CDS Astronomical Databases SIMBAD and VIZIER, the Wide-field Infrared Survey Explorer, and the Two Micron All Sky Survey. 

\facilities{APO:3.5m, CHARA, CTIO:1.5m, TSU:AST} 

\software{ Grid Search for Binary Stars \citep{schaefer16}, MESA \citep{mesa1}, RVFIT \citep{rvfit},  SearchCal \citep{searchcal}, TODCOR \citep{todcor1}, Yonsei-Yale models \citep{y2}}

\clearpage

% -----------------------------------------------------------------------------
\appendix

\input{table_rv.txt}

\input{table_cal.txt}

\break

\begin{figure*}
\centering
\includegraphics[width=0.9\textwidth]{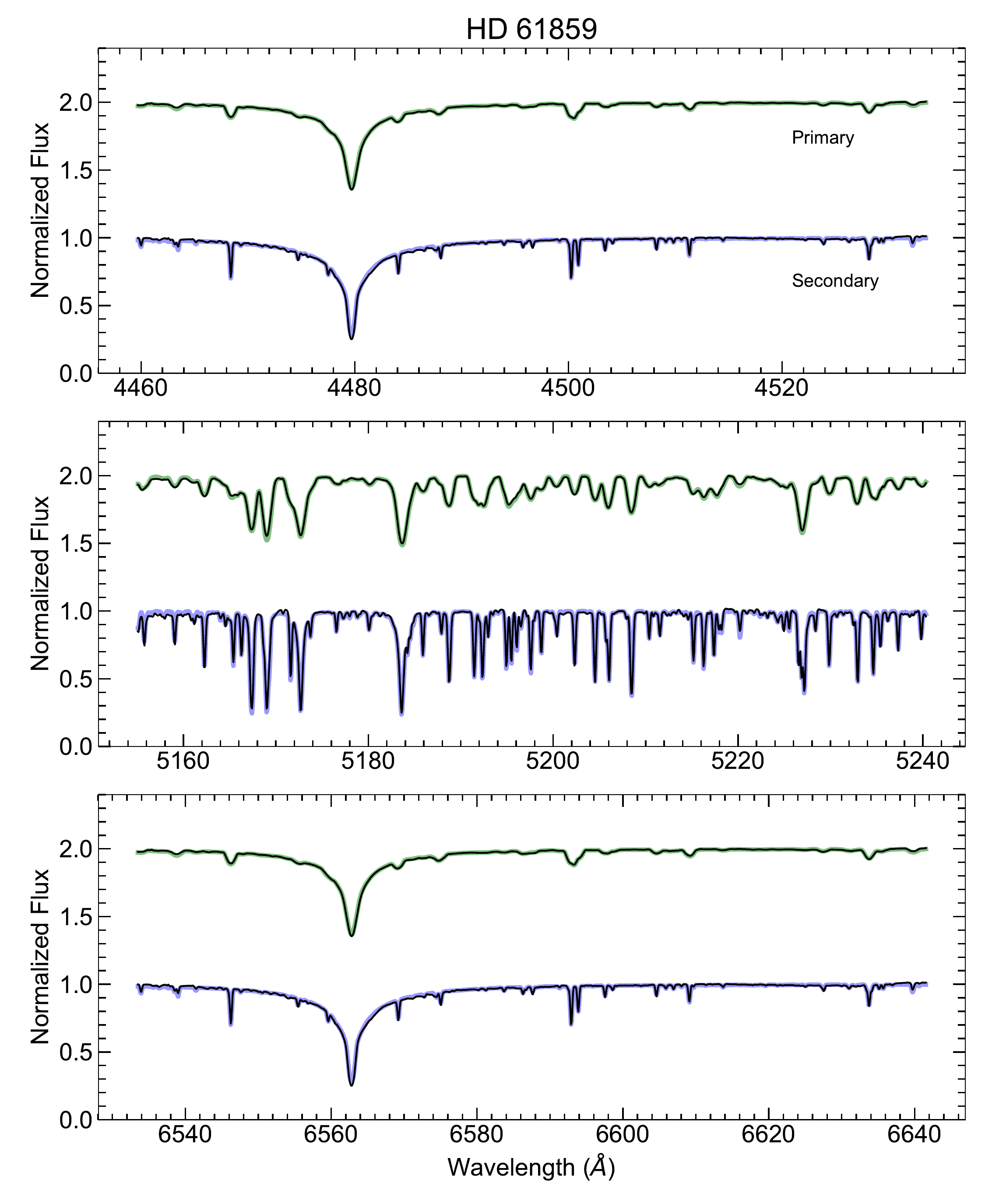}
\caption{Reconstructed spectra of HD~61859 around H$\beta$ (top), the Mg~b triplet (middle), and H$\alpha$ (bottom), for example. The reconstructed spectra are shown in black, and the best-fit model spectra are overplotted in green for the primary and blue for the secondary. 
\label{rec61859}}
\end{figure*}

\begin{figure*}
\centering
\includegraphics[width=0.9\textwidth]{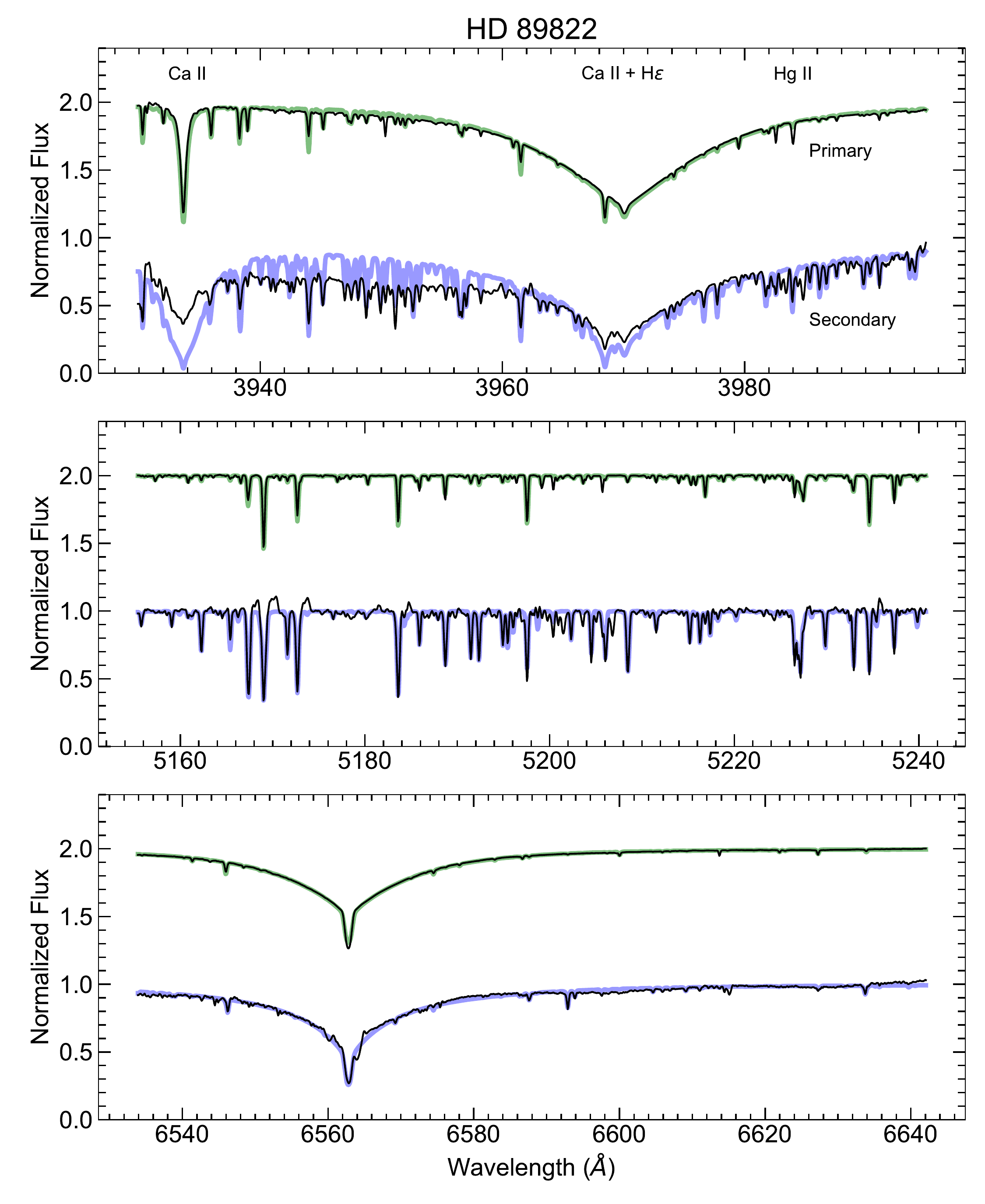}
\caption{Reconstructed spectra of HD~89822 around the Ca H \& K lines (top), the Mg~b triplet (middle), and H$\alpha$ (bottom), for example. The reconstructed spectra are shown in black, and the best-fit model spectra are overplotted in green for the primary and blue for the secondary. In the top panel, lines relevant to HgMn and Am stars are labeled, including Ca~K 3934\AA, a blend of Ca~H 3969\AA\ with H$\epsilon$ 3970\AA, and Hg II 3984\AA. The Hg II line is much stronger in the primary star than in the models as expected for HgMn stars, and the Ca~K line of the secondary component is much weaker than in the models as expected for metallic line stars. 
\label{rec89822}}
\end{figure*}

\begin{figure*}
\centering
\includegraphics[width=0.9\textwidth]{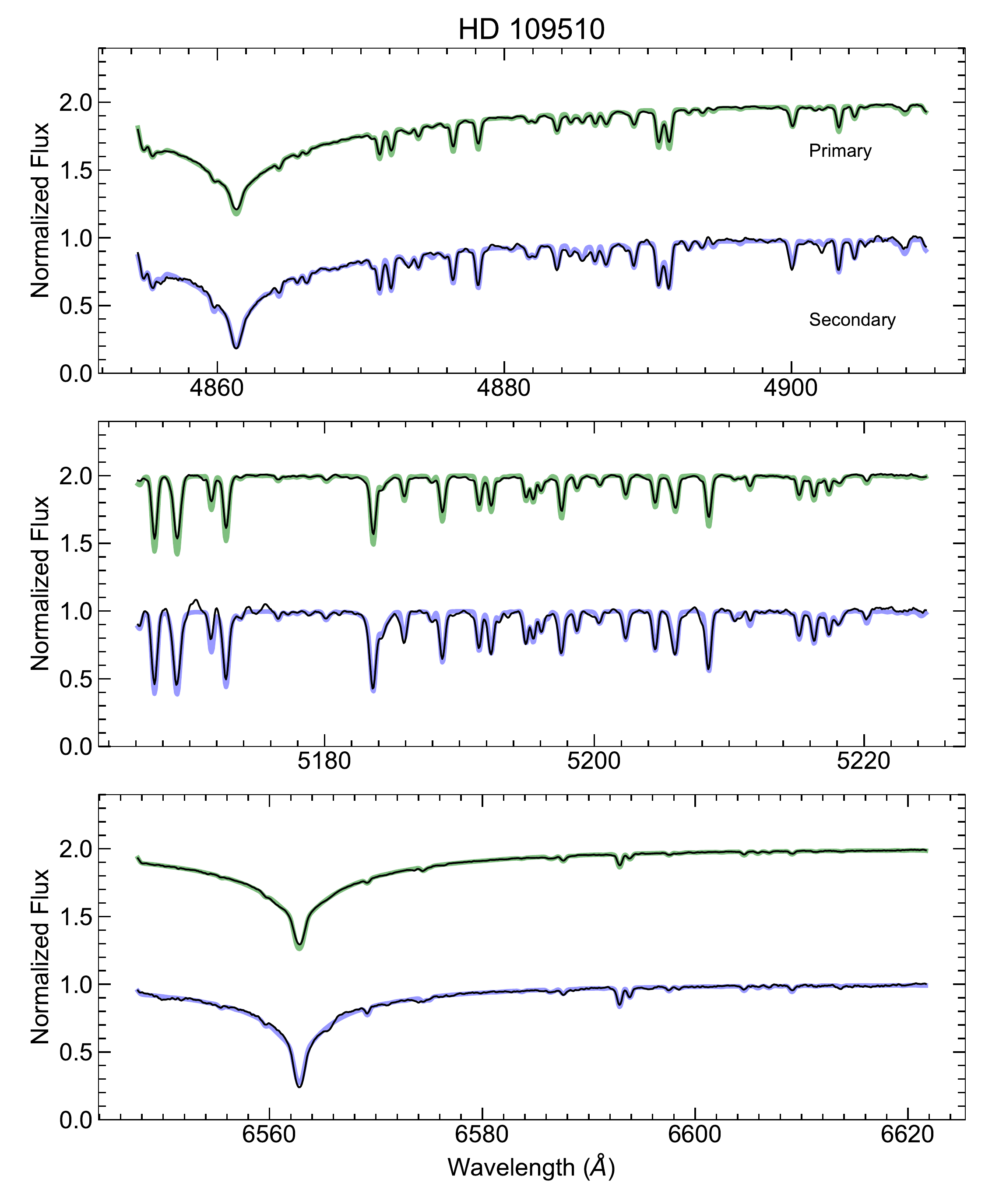}
\caption{Reconstructed spectra of HD~109510 around H$\beta$ (top), the Mg~b triplet (middle), and H$\alpha$ (bottom), for example. The reconstructed spectra are shown in black, and the best-fit model spectra are overplotted in green for the primary and blue for the secondary.
\label{rec109510}}
\end{figure*}

\begin{figure*}
\centering
\includegraphics[width=0.9\textwidth]{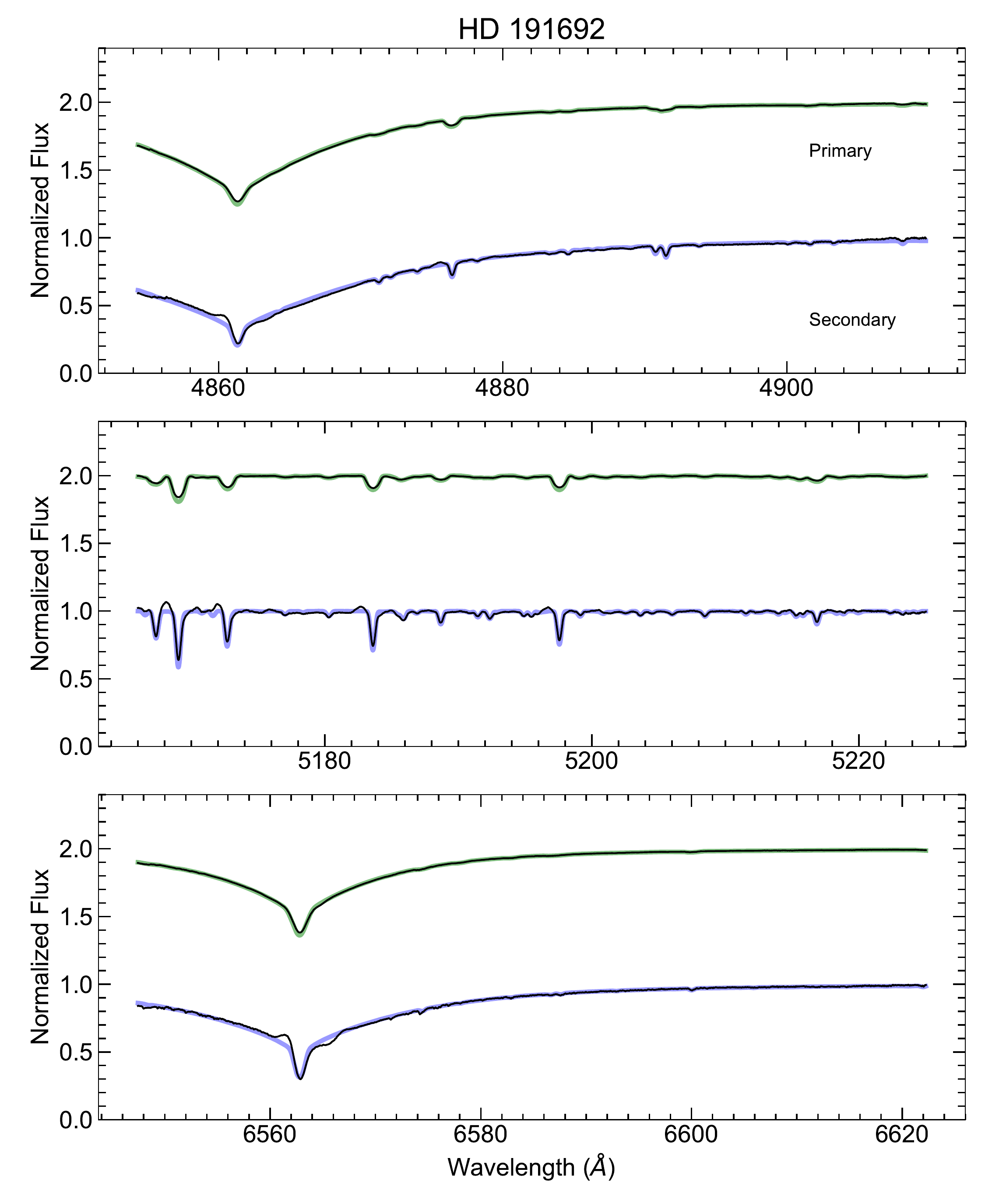}
\caption{Reconstructed spectra of HD~191692 around H$\beta$ (top), the Mg~b triplet (middle), and H$\alpha$ (bottom), for example. The reconstructed spectra are shown in black, and the best-fit model spectra are overplotted in green for the primary and blue for the secondary. 
\label{rec191692}}
\end{figure*}

\clearpage

% -----------------------------------------------------------------------------

\end{document}

%% file: table_relpos.txt
%\startlongtable
%\centering
\begin{deluxetable*}{llcccccccc}
\tablewidth{0pt}
\tabletypesize{\normalsize}
\tablecaption{Relative Positions  \vspace{6pt} \label{relpos}}
\tablehead{ \colhead{Target} & \colhead{UT Date} & \colhead{HJD-2,400,000} & \colhead{$\rho$} 
& \colhead{$\theta$} & \colhead{$\sigma_{maj}$} & \colhead{$\sigma_{min}$} & \colhead{$\phi$} & \colhead{$f_2/f_1$} &  \colhead{Beam}\\
\colhead{ } & \colhead{ } & \colhead{ } & \colhead{(mas)} & \colhead{(deg)} &  \colhead{(mas)}  
&  \colhead{(mas)}  &  \colhead{(deg)}  &  \colhead{ } &  \colhead{Combiner}  }
\startdata 
%     &    CLIMB     \\
\hline
HD 61859 & 2013 Dec 11   &  56638.0297    &   2.531   &   63.3   &   0.253   &   0.056   &  83.6  &  0.410 $\pm$  0.003        & CLIMB  \\
HD 61859 & 2017 Feb 01   &  57785.7409    &   3.719   &  242.1   &   0.169   &   0.101   &  51.7  &  0.406 $\pm$  0.009        & CLIMB  \\
HD 61859 & 2017 Nov 30   &  58087.8801    &   2.961   &   60.5   &   0.171   &   0.092   & 130.9  &  0.390 $\pm$  0.005        & CLIMB  \\
HD 61859 & 2018 Dec 12   &  58464.8835    &   2.494   &   63.4   &   0.171   &   0.083   &  56.2  &  0.431 $\pm$  0.014        & CLIMB  \\
HD 61859 & 2019 Sep 18   &  58745.0249    &   0.430   &  140.0   &   0.226   &   0.226   & 144.5  &  0.574 $\pm$  0.180        & CLIMB  \\
HD 61859 & 2019 Dec 21   &  58838.8581    &   0.696   &  194.7   &   0.270   &   0.115   &  57.8  &  \nodata                   & CLIMB  \\
HD 61859 & 2021 Mar 28   &  59301.6728    &   5.027   &  237.4   &   0.001   &   0.001   & 153.4  &  0.3852 $\pm$  0.0002      & MIRC-X   \\
\hline
HD 89822 & 2017 Nov 30   &  58088.0567    &   1.884   &  282.1   &   0.131   &   0.104   &    28.3   &   0.214 $\pm$ 0.008     & CLIMB  \\
HD 89822 & 2018 Apr 10   &  58218.7122    &   1.002   &  166.8   &   0.110   &   0.027   &   164.4   &   0.232 $\pm$ 0.002     & CLIMB  \\
HD 89822 & 2018 Apr 11   &  58219.7138    &   1.128   &  122.2   &   0.115   &   0.058   &   100.1   &   0.231 $\pm$ 0.004     & CLIMB  \\
HD 89822 & 2018 Nov 26   &  58449.0307    &   1.206   &  227.7   &   0.063   &   0.049   &   153.5   &   0.263 $\pm$ 0.005     & CLIMB  \\
HD 89822 & 2018 Dec 12   &  58464.9750    &   1.315   &   22.7   &   0.069   &   0.059   &   150.6   &   0.267 $\pm$ 0.005     & CLIMB  \\
HD 89822 & 2019 Apr 25   &  58598.7623    &   1.386   &  259.3   &   0.067   &   0.034   &   110.4   &   0.339 $\pm$ 0.011     & CLIMB  \\
HD 89822 & 2019 Apr 26   &  58599.6880    &   1.207   &  224.8   &   0.082   &   0.050   &    43.0   &   0.396 $\pm$ 0.030     & CLIMB  \\
HD 89822 & 2019 Apr 27   &  58600.7010    &   1.060   &  170.9   &   0.114   &   0.058   &   138.3   &   0.245 $\pm$ 0.004     & CLIMB  \\
HD 89822 & 2019 Apr 28   &  58601.7970    &   1.087   &  115.6   &   0.219   &   0.087   &   115.5   &   0.278 $\pm$ 0.007     & CLIMB  \\
HD 89822 & 2019 Dec 20   &  58838.0461    &   2.071   &  322.4   &   0.087   &   0.052   &    72.0   &   0.208 $\pm$ 0.003     & CLIMB  \\
HD 89822 & 2021 Mar 28   &  59301.7480    &   2.106   &  319.5   &   0.001   &   0.001   &   143.1   &   0.2276 $\pm$ 0.0001   & MIRC-X   \\
\hline
HD 109510 & 2017 May 20   &  57893.7031   &   0.680   &  167.6   &   0.267   &   0.090   &  161.4   &   0.471 $\pm$  0.102    &  CLIMB  \\
HD 109510 & 2017 May 21   &  57894.7054   &   0.587   &  285.4   &   0.134   &   0.066   &  111.5   &   0.458 $\pm$  0.065    &  CLIMB  \\
HD 109510 & 2019 Apr 26   &  58599.8104   &   0.831   &  319.7   &   0.027   &   0.014   &  136.6   &   0.561 $\pm$  0.011    &  CLIMB  \\
HD 109510 & 2019 Apr 27   &  58600.8133   &   0.689   &  354.2   &   0.041   &   0.021   &  102.5   &   0.579 $\pm$  0.015    &  CLIMB  \\
HD 109510 & 2021 Mar 28   &  59301.8264   &   0.935   &  146.8   &   0.001   &   0.001   &  168.7   &   0.5285 $\pm$ 0.0002   &  MIRC-X   \\
\hline
HD 191692 & 2012 Sep 04   &  56174.6984   &   1.375   &  296.1   &   0.383   &   0.240   &  130.9   &   0.181 $\pm$  0.012    & CLIMB  \\
HD 191692 & 2017 Sep 08   &  58004.6859   &   2.795   &   17.7   &   0.236   &   0.236   &  121.0   &   0.235 $\pm$  0.001    & CLIMB  \\
HD 191692 & 2018 Aug 17   &  58347.8136   &   2.308   &    4.8   &   0.248   &   0.136   &   15.9   &   0.260 $\pm$  0.003    & CLIMB  \\
HD 191692 & 2018 Sep 03   &  58364.6972   &   2.439   &    9.8   &   0.201   &   0.178   &  112.6   &   0.245 $\pm$  0.004    & CLIMB  \\
HD 191692 & 2018 Sep 05   &  58366.6994   &   1.337   &  287.9   &   0.293   &   0.188   &  141.2   &   0.169 $\pm$  0.002    & CLIMB  \\
HD 191692 & 2019 Jul 04   &  58668.9676   &   4.383   &   55.8   &   0.007   &   0.002   &   50.2   &   0.238 $\pm$  0.001    & MIRC-X  \\
HD 191692 & 2019 Sep 16   &  58742.6754   &   1.701   &  330.8   &   0.496   &   0.252   &   41.1   &   0.177 $\pm$  0.004    & CLIMB  \\
HD 191692 & 2020 Aug 12   &  59073.7005   &   4.111   &   95.3   &   0.006   &   0.002   &  141.9   &   0.2415 $\pm$  0.0001  & MIRC-X  \\
HD 191692 & 2020 Aug 13   &  59074.6859   &   4.476   &   88.2   &   0.010   &   0.004   &  147.9   &   0.2235 $\pm$  0.0003  & MIRC-X  \\
\enddata  
\end{deluxetable*} 

%% file: table_orbpar.txt
\begin{deluxetable*}{lcccc}
\tablewidth{0pt}
\tabletypesize{\normalsize}
\tablecaption{Orbital Parameters from VB+SB2 Solution \vspace{6pt} \label{orbpar}}
\tablehead{
\colhead{Parameter}       & \colhead{HD 61859}          & \colhead{HD 89822}         & \colhead{HD 109510}         & \colhead{HD 191692}     } 
\startdata	                                              
$P$ (days)	          &  $31.500002\pm0.000056$    &	$11.579113\pm0.000010$	  &	$7.336673\pm0.000087$	  &	$17.124281\pm0.000038$  \\
$T$ (HJD-2400000)	  &  $58880.701\pm0.018   $    &	$57756.168\pm0.005   $	  &	$59361.967\pm0.011  $	  &	$58624.154\pm0.004   $  \\
$e$			          &  $0.1951 \pm 0.0007   $    &	$0.2943  \pm 0.0009  $	  &	$0.2585	\pm 0.0012  $	  &	$0.6040	\pm 0.0009   $  \\
$\omega_1$ (deg)	  &  $39.52  \pm 0.23     $    &	$176.50  \pm 0.20    $	  &	$302.33	\pm 0.28    $	  &	$33.73	\pm 0.22     $  \\
$i$ (deg)		      &  $94.73  \pm 3.51     $    &	$141.87  \pm 0.97    $	  &	$61.40	\pm 3.89    $	  &	$144.10	\pm 0.18     $  \\
$\alpha$ (mas)		  &  $4.403  \pm 0.003    $    &	$1.634   \pm 0.001   $	  &	$1.007	\pm 0.037   $	  &	$3.148	\pm 0.002    $  \\
$\Omega$ (deg)		  &  $237.81 \pm 3.45     $    &	$133.49  \pm 0.13    $	  &	$136.17	\pm 2.86    $	  &	$96.92	\pm 0.19     $  \\
$\gamma$ (km s$^{-1}$)&  $-12.51 \pm 0.03     $    &	$-2.24   \pm 0.03    $	  &	$3.51	\pm 0.06    $	  &	$-29.26	\pm 0.08     $  \\
$K_1$ (km s$^{-1}$)	  &  $46.77  \pm 0.07     $    &	$38.17   \pm 0.04    $	  &	$68.16	\pm 0.09    $	  &	$48.78	\pm 0.09     $  \\
$K_2$ (km s$^{-1}$)	  &  $52.81  \pm 0.03     $    &	$62.11   \pm 0.09    $	  &	$81.28	\pm 0.20    $	  &	$63.48	\pm 0.11     $  \\
\enddata                   
\end{deluxetable*}

%% file: table_atmospar.txt
\begin{deluxetable*}{lcccc}
\tablewidth{0pt}
\tabletypesize{\normalsize}
\tablecaption{Astrophysical Parameters  \vspace{6pt} \label{atmospar}}
\tablehead{
\colhead{Parameter} &\colhead{ \ \ \ HD 61859 \ \ \ } & \colhead{ \ \ \ HD 89822 \ \ \ } & \colhead{ \ \ \ HD 109510 \ \ \ } & \colhead{ \ \ \ HD 191692 \ \ \ }}
\startdata		%             HD 61859	           HD 89822              HD 109510 	      HD 191692 	       
$M_1$ ($M_\odot$)	     &   $1.629\pm0.023$     & $2.779\pm0.153$   & $1.838\pm0.218$   & $3.564\pm0.049$    \\[2pt] 
$M_2$ ($M_\odot$)	     &   $1.443\pm0.020$     & $1.708\pm0.094$   & $1.541\pm0.184$   & $2.739\pm0.037$    \\[2pt] 
$R_1$ ($R_\odot$)	     &   $2.53\pm0.10$       & $3.16\pm0.11$     & $2.28\pm0.06$     & $4.76\pm0.14$	  \\[2pt] 
$R_2$ ($R_\odot$)	     &   $1.51\pm0.06$       & $1.73\pm0.06$     & $1.70\pm0.07$     & $2.34\pm0.07$	  \\[2pt] 
$T_{\rm eff \ 1}$ (K)    &   $6390\pm180$        & $10260\pm100$     & $7630\pm120$      & $10300\pm200$	  \\[2pt] 
$T_{\rm eff \ 2}$ (K)    &   $6610\pm230$        & $ 7860\pm140$     & $7180\pm140$      & $10230\pm220$	  \\[2pt] 
$\log g_1$ (cgs)	     &   $3.85\pm0.02$       & $3.88\pm0.05$     & $3.86\pm0.01$     & $3.64\pm0.02$	  \\[2pt] 
$\log g_2$ (cgs)	     &   $4.22\pm0.02$       & $4.22\pm0.05$     & $4.09\pm0.02$     & $4.14\pm0.02$	  \\[2pt]
$L_1$ ($L_\odot$)	     &   $9.6\pm1.3$         & $101.0\pm8.0$     & $15.6\pm1.3$      & $229.8\pm22.5$     \\[2pt]
$L_2$ ($L_\odot$)	     &   $4.1\pm0.5$         & $  9.7\pm1.0$     & $ 7.2\pm0.9$      & $ 54.0\pm 5.8$     \\[2pt]
$V_1 \sin i$ (\kms)	     &   $37.1\pm1.0$        & $\le 4.2$         & $14.5\pm1.5$      & $36.3\pm0.4$	      \\[2pt] 
$V_2 \sin i$ (\kms)	     &   $\le 4.2$           & $5.1\pm2.1$       & $14.2\pm1.1$      & $13.6\pm0.9$	      \\[2pt] 
$\log Z_1/Z_\odot$ (dex) &   $-0.05\pm0.12$      & $ 0.11\pm0.04$    & $-0.05\pm0.02$    & $0.05\pm0.05$	  \\[2pt] 
$\log Z_2/Z_\odot$ (dex) &   $-0.06\pm0.05$      & $-0.05\pm0.07$    & $-0.04\pm0.07$    & $0.12\pm0.13$	  \\[2pt] 
Distance (pc, this work) &   $64.4\pm0.3$        & $101.1\pm2.0$     & $110.1\pm6.1$     & $76.3\pm0.3$	      \\[2pt]
Distance (pc, Gaia DR3)	 &   $64.8\pm1.9$        & $103.8\pm1.0$     & $111.8\pm0.4$     & $70.1\pm2.3$	      \\[2pt]
\enddata     
\vspace{12pt}              
\end{deluxetable*}

%% file: table_sed.txt
\begin{deluxetable*}{lcccc}
\tablewidth{0pt}
\tabletypesize{\normalsize}
\tablecaption{SED Parameters  \vspace{6pt} \label{sedpar}}
\tablehead{
\colhead{Parameter} &\colhead{ \ \ \ HD 61859 \ \ \ } & \colhead{ \ \ \ HD 89822 \ \ \ } & \colhead{ \ \ \ HD 109510 \ \ \ } & \colhead{ \ \ \ HD 191692 \ \ \ }}
\startdata		%             HD 61859	           HD 89822              HD 109510 	           HD 191692 	       
$f_2/f_1$ (6500\AA)	&   $0.45\pm0.03$       &   $0.18\pm0.07$       &   $0.70\pm0.05$       &   $0.25\pm0.06$       \\[2pt] 
$f_2/f_1$ ($H$-band)&   $0.385\pm0.007$     &   $0.228\pm0.007$     &   $0.526\pm0.007$     &   $0.238\pm0.004$     \\[2pt] 
$R_2/R_1$ 	   		&   $0.60\pm0.01$       &   $0.55\pm0.01$       &   $0.75\pm0.02$       &   $0.49\pm0.01$       \\[2pt] 
$\theta_1$ (mas)    &   $0.363\pm0.014$     &   $0.291\pm0.009$     &   $0.189\pm0.005$     &   $0.580\pm0.017$\tablenotemark{*}  \\[2pt] 
$\theta_2$ (mas)    &   $0.217\pm0.009$     &   $0.159\pm0.005$     &   $0.142\pm0.007$     &   $0.285\pm0.009$     \\[2pt] 
$E(B-V)$    		&   $0.05\pm0.01$       &   $0.01\pm0.02$       &   $0.01\pm0.01$       &   $0.00\pm0.01$       \\[2pt]
\enddata     
\tablenotetext{*}{Fixed to the limb-darkened angular diameter from CHARA observations}
\vspace{12pt}              
\end{deluxetable*}

%% file: table_rv.txt
\startlongtable
\begin{deluxetable*}{llcrcrcc}	
\tablewidth{0pt}
\tabletypesize{\footnotesize}
\tablecaption{  Radial Velocity Measurements \vspace{6pt} \label{rvtable}    }
\tablehead{ 
\colhead{Target} & \colhead{UT Date} & \colhead{HJD-2,400,000} & \colhead{$V_{r\ 1}$} & \colhead{$\sigma_1$} 
&  \colhead{$V_{r\ 2}$} & \colhead{$\sigma_2$} & \colhead{Source}   \\  
\colhead{} &\colhead{} & \colhead{} & \colhead{(km~s$^{-1}$)} & \colhead{(km~s$^{-1}$)} 
& \colhead{(km~s$^{-1}$)} & \colhead{(km~s$^{-1}$)} & }
\startdata
%    DATE	HJD		RV1		ERV1		RV2		ERV2		Source
\hline
HD 61859  &  2011 Oct 06  & 55840.9727   &  $-41.92$    &    0.52   &   $ 20.58$  &  0.26  &    Fairborn \\[-4pt] 
HD 61859  &  2011 Oct 15  & 55849.9297   &  $ 18.28$    &    0.52   &   $-46.32$  &  0.26  &    Fairborn \\[-4pt] 
HD 61859  &  2011 Nov 23  & 55888.7539   &  $ 24.98$    &    0.52   &   $-55.02$  &  0.26  &    Fairborn \\[-4pt] 
HD 61859  &  2012 Jan 01  & 55927.7266   &  $-47.12$    &    0.52   &   $ 26.38$  &  0.26  &    Fairborn \\[-4pt] 
HD 61859  &  2012 Jan 18  & 55944.8945   &  $ 21.68$    &    0.52   &   $-50.92$  &  0.26  &    Fairborn \\[-4pt] 
HD 61859  &  2012 Feb 25  & 55982.8242   &  $ 29.28$    &    0.52   &   $-59.92$  &  0.26  &    Fairborn \\[-4pt] 
HD 61859  &  2012 Apr 04  & 56021.6875   &  $-45.62$    &    0.52   &   $ 23.78$  &  0.26  &    Fairborn \\[-4pt] 
HD 61859  &  2012 May 13  & 56060.6953   &  $-44.72$    &    0.52   &   $ 23.58$  &  0.26  &    Fairborn \\[-4pt] 
HD 61859  &  2012 Oct 16  & 56216.8750   &  $-48.32$    &    0.52   &   $ 28.58$  &  0.26  &    Fairborn \\[-4pt] 
HD 61859  &  2012 Nov 04  & 56235.8281   &  $ 19.58$    &    0.52   &   $-47.82$  &  0.26  &    Fairborn \\[-4pt] 
HD 61859  &  2012 Dec 17  & 56278.7109   &  $-51.42$    &    0.52   &   $ 30.68$  &  0.26  &    Fairborn \\[-4pt] 
HD 61859  &  2013 Jan 05  & 56297.9297   &  $ 28.18$    &    0.52   &   $-58.92$  &  0.26  &    Fairborn \\[-4pt] 
HD 61859  &  2013 Jan 16  & 56308.9297   &  $-52.02$    &    0.52   &   $ 32.48$  &  0.26  &    Fairborn \\[-4pt] 
HD 61859  &  2013 Feb 01  & 56324.9414   &  $ 35.28$    &    0.52   &   $-66.92$  &  0.26  &    Fairborn \\[-4pt] 
HD 61859  &  2013 Feb 17  & 56340.8398   &  $-51.72$    &    0.52   &   $ 32.18$  &  0.26  &    Fairborn \\[-4pt] 
HD 61859  &  2013 Mar 24  & 56375.8242   &  $-45.12$    &    0.52   &   $ 23.18$  &  0.26  &    Fairborn \\[-4pt] 
HD 61859  &  2013 Sep 21  & 56557.0156   &  $-44.82$    &    0.52   &   $ 23.38$  &  0.26  &    Fairborn \\[-4pt] 
HD 61859  &  2013 Oct 11  & 56576.9414   &  $ 36.18$    &    0.52   &   $-66.42$  &  0.26  &    Fairborn \\[-4pt] 
HD 61859  &  2013 Nov 26  & 56622.8594   &  $-51.42$    &    0.52   &   $ 32.28$  &  0.26  &    Fairborn \\[-4pt] 
HD 61859  &  2013 Dec 16  & 56642.9688   &  $ 39.18$    &    0.52   &   $-70.42$  &  0.26  &    Fairborn \\[-4pt] 
HD 61859  &  2014 Jan 01  & 56659.0195   &  $-45.12$    &    0.52   &   $ 24.48$  &  0.26  &    Fairborn \\[-4pt] 
HD 61859  &  2014 Jan 17  & 56674.9531   &  $ 36.18$    &    0.52   &   $-67.62$  &  0.26  &    Fairborn \\[-4pt] 
HD 61859  &  2014 Mar 17  & 56733.8281   &  $ 31.18$    &    0.52   &   $-62.52$  &  0.26  &    Fairborn \\[-4pt] 
HD 61859  &  2014 Oct 28  & 56958.9648   &  $ 32.78$    &    0.52   &   $-63.02$  &  0.26  &    Fairborn \\[-4pt] 
HD 61859  &  2014 Dec 30  & 57021.9961   &  $ 33.08$    &    0.52   &   $-63.02$  &  0.26  &    Fairborn \\[-4pt] 
HD 61859  &  2015 Feb 06  & 57059.9023   &  $-37.52$    &    0.52   &   $ 15.68$  &  0.26  &    Fairborn \\[-4pt] 
HD 61859  &  2015 Oct 03  & 57298.9531   &  $ 18.88$    &    0.52   &   $-46.62$  &  0.26  &    Fairborn \\[-4pt] 
HD 61859  &  2015 Nov 21  & 57347.8672   &  $-52.42$    &    0.52   &   $ 32.68$  &  0.26  &    Fairborn \\[-4pt] 
HD 61859  &  2016 Jan 26  & 57413.6484   &  $-48.41$    &    0.49   &   $ 28.79$  &  0.34  &       APO \\[-4pt] 
HD 61859  &  2016 Feb 11  & 57429.8359   &  $ 40.78$    &    0.52   &   $-73.32$  &  0.26  &    Fairborn \\[-4pt] 
HD 61859  &  2016 Mar 15  & 57462.8203   &  $ 33.78$    &    0.52   &   $-65.42$  &  0.26  &    Fairborn \\[-4pt] 
HD 61859  &  2016 Apr 14  & 57492.7852   &  $ 41.18$    &    0.52   &   $-73.32$  &  0.26  &    Fairborn \\[-4pt] 
HD 61859  &  2016 May 10  & 57518.6641   &  $ 12.28$    &    0.52   &   $-39.42$  &  0.26  &    Fairborn \\[-4pt] 
HD 61859  &  2016 Sep 16  & 57648.0078   &  $ 36.18$    &    0.52   &   $-67.12$  &  0.26  &    Fairborn \\[-4pt] 
HD 61859  &  2016 Oct 16  & 57678.0000   &  $ 26.18$    &    0.52   &   $-55.72$  &  0.26  &    Fairborn \\[-4pt] 
HD 61859  &  2017 Jan 26  & 57779.8008   &  $ 13.48$    &    0.52   &   $-40.82$  &  0.26  &    Fairborn \\[-4pt] 
HD 61859  &  2017 Mar 09  & 57821.7266   &  $-51.07$    &    0.47   &   $ 31.64$  &  0.34  &       APO \\[-4pt] 
HD 61859  &  2017 Mar 13  & 57825.7891   &  $-41.22$    &    0.52   &   $ 18.28$  &  0.26  &    Fairborn \\[-4pt] 
HD 61859  &  2017 Apr 04  & 57847.6133   &  $-38.00$    &    0.46   &   $ 17.10$  &  0.33  &       APO \\[-4pt] 
HD 61859  &  2017 Apr 13  & 57856.7305   &  $-42.32$    &    0.52   &   $ 21.28$  &  0.26  &    Fairborn \\[-4pt] 
HD 61859  &  2017 May 27  & 57900.6562   &  $ 39.48$    &    0.52   &   $-70.92$  &  0.26  &    Fairborn \\[-4pt] 
HD 61859  &  2017 Sep 18  & 58014.9219   &  $-40.72$    &    0.52   &   $ 18.18$  &  0.26  &    Fairborn \\[-4pt] 
HD 61859  &  2017 Oct 30  & 58056.8828   &  $ 31.88$    &    0.52   &   $-62.92$  &  0.26  &    Fairborn \\[-4pt] 
HD 61859  &  2017 Dec 02  & 58089.8438   &  $ 38.59$    &    0.48   &   $-71.43$  &  0.34  &       APO \\[-4pt] 
HD 61859  &  2017 Dec 12  & 58099.8633   &  $-40.52$    &    0.52   &   $ 19.08$  &  0.26  &    Fairborn \\[-4pt] 
HD 61859  &  2018 Jan 04  & 58122.6953   &  $ 40.39$    &    0.49   &   $-72.88$  &  0.35  &       APO \\[-4pt] 
HD 61859  &  2018 Mar 19  & 58196.8555   &  $-51.12$    &    0.52   &   $ 30.08$  &  0.26  &    Fairborn \\[-4pt] 
HD 61859  &  2018 Apr 04  & 58212.7109   &  $ 18.65$    &    0.45   &   $-49.94$  &  0.32  &       APO \\[-4pt] 
HD 61859  &  2018 May 13  & 58251.7070   &  $ 19.78$    &    0.52   &   $-49.82$  &  0.26  &    Fairborn \\[-4pt] 
HD 61859  &  2018 Oct 29  & 58420.9922   &  $-50.82$    &    0.52   &   $ 30.48$  &  0.26  &    Fairborn \\[-4pt] 
HD 61859  &  2018 Nov 16  & 58438.7891   &  $ 36.05$    &    0.47   &   $-69.16$  &  0.34  &       APO \\[-4pt] 
HD 61859  &  2018 Dec 24  & 58476.8945   &  $-34.22$    &    0.52   &   $ 11.18$  &  0.26  &    Fairborn \\[-4pt] 
HD 61859  &  2019 Jan 14  & 58497.6406   &  $ 29.80$    &    0.48   &   $-61.33$  &  0.35  &       APO \\[-4pt] 
HD 61859  &  2019 Jan 15  & 58498.7422   &  $ 36.86$    &    1.00   &   $-67.55$  &  0.72  &       APO \\[-4pt] 
HD 61859  &  2019 Jan 19  & 58502.8945   &  $ 28.35$    &    0.44   &   $-58.67$  &  0.31  &       APO \\[-4pt] 
HD 61859  &  2019 Jan 31  & 58514.9023   &  $-51.08$    &    0.48   &   $ 30.98$  &  0.34  &       APO \\[-4pt] 
HD 61859  &  2019 Feb 25  & 58539.8281   &  $-33.82$    &    0.52   &   $ 10.18$  &  0.26  &    Fairborn \\[-4pt] 
HD 61859  &  2019 Mar 24  & 58566.6914   &  $ 19.95$    &    0.44   &   $-49.33$  &  0.32  &       APO \\[-4pt] 
HD 61859  &  2019 Oct 14  & 58770.9414   &  $-39.27$    &    0.48   &   $ 17.66$  &  0.35  &       APO \\[-4pt] 
HD 61859  &  2019 Nov 13  & 58800.8242   &  $-46.72$    &    0.52   &   $ 25.08$  &  0.26  &    Fairborn \\[-4pt] 
HD 61859  &  2019 Nov 14  & 58801.8203   &  $-43.22$    &    0.52   &   $ 20.88$  &  0.26  &    Fairborn \\[-4pt] 
HD 61859  &  2019 Nov 14  & 58802.0078   &  $-40.39$    &    0.47   &   $ 20.32$  &  0.34  &       APO \\[-4pt] 
HD 61859  &  2020 Jan 12  & 58860.7578   &  $-50.79$    &    0.46   &   $ 32.45$  &  0.32  &       APO \\[-4pt] 
HD 61859  &  2020 Jan 12  & 58860.7891   &  $-52.12$    &    0.52   &   $ 32.08$  &  0.26  &    Fairborn \\[-4pt] 
HD 61859  &  2020 Jan 27  & 58875.6953   &  $ 30.88$    &    0.52   &   $-61.62$  &  0.26  &    Fairborn \\[-4pt] 
HD 61859  &  2020 Jan 28  & 58876.7081   &  $ 37.08$    &    0.52   &   $-68.22$  &  0.26  &    Fairborn \\[-4pt] 
HD 61859  &  2020 Jan 29  & 58877.7080   &  $ 41.58$    &    0.52   &   $-72.62$  &  0.26  &    Fairborn \\[-4pt] 
HD 61859  &  2020 Jan 31  & 58879.7079   &  $ 37.98$    &    0.52   &   $-69.22$  &  0.26  &    Fairborn \\[-4pt] 
HD 61859  &  2020 Feb 01  & 58880.7079   &  $ 30.08$    &    0.52   &   $-61.42$  &  0.26  &    Fairborn \\[-4pt] 
HD 61859  &  2020 Feb 02  & 58881.7079   &  $ 20.68$    &    0.52   &   $-49.32$  &  0.26  &    Fairborn \\[-4pt] 
HD 61859  &  2020 Feb 07  & 58886.7077   &  $-36.22$    &    0.52   &   $ 13.48$  &  0.26  &    Fairborn \\[-4pt] 
HD 61859  &  2020 Feb 08  & 58887.7077   &  $-42.52$    &    0.52   &   $ 21.18$  &  0.26  &    Fairborn \\[-4pt] 
HD 61859  &  2020 Feb 09  & 58888.7076   &  $-47.02$    &    0.52   &   $ 26.28$  &  0.26  &    Fairborn \\[-4pt] 
HD 61859  &  2020 Feb 14  & 58893.7074   &  $-49.92$    &    0.52   &   $ 29.98$  &  0.26  &    Fairborn \\[-4pt] 
HD 61859  &  2020 Feb 14  & 58893.7539   &  $-49.11$    &    0.46   &   $ 29.98$  &  0.32  &       APO \\ 
\hline
HD 89822  &  2005 Feb 09  & 53410.8867   &  $ 12.20$  &    0.22  &   $ -25.17$  &    0.49  &  Fairborn \\[-4pt] 
HD 89822  &  2005 Apr 01  & 53461.8945   &  $-21.91$  &    0.22  &   $  29.61$  &    0.49  &  Fairborn \\[-4pt] 
HD 89822  &  2005 Apr 16  & 53476.8242   &  $ 22.87$  &    0.22  &   $ -43.30$  &    0.49  &  Fairborn \\[-4pt] 
HD 89822  &  2005 May 05  & 53495.9180   &  $-39.67$  &    0.22  &   $  60.16$  &    0.49  &  Fairborn \\[-4pt] 
HD 89822  &  2005 May 21  & 53511.8711   &  $ 23.34$  &    0.22  &   $ -43.76$  &    0.49  &  Fairborn \\[-4pt] 
HD 89822  &  2006 Jan 31  & 53766.8867   &  $ 24.40$  &    0.22  &   $ -46.04$  &    0.49  &  Fairborn \\[-4pt] 
HD 89822  &  2006 Apr 17  & 53842.8672   &  $-48.70$  &    0.22  &   $  73.82$  &    0.49  &  Fairborn \\[-4pt] 
HD 89822  &  2006 May 14  & 53869.8242   &  $ 19.00$  &    0.22  &   $ -36.81$  &    0.49  &  Fairborn \\[-4pt] 
HD 89822  &  2006 May 30  & 53885.7500   &  $ 10.52$  &    0.22  &   $ -21.62$  &    0.49  &  Fairborn \\[-4pt] 
HD 89822  &  2017 Jan 11  & 57764.8320   &  $  9.09$  &    0.53  &   $ -20.18$  &    1.54  &     APO \\[-4pt] 
HD 89822  &  2017 Feb 16  & 57800.8867   &  $-16.10$  &    0.51  &   $  22.17$  &    1.56  &     APO \\[-4pt] 
HD 89822  &  2017 Mar 09  & 57821.7422   &  $ 18.95$  &    0.55  &   $ -36.05$  &    1.64  &     APO \\[-4pt] 
HD 89822  &  2017 Apr 04  & 57847.6406   &  $-28.27$  &    0.54  &   $  40.76$  &    1.62  &     APO \\[-4pt] 
HD 89822  &  2017 Apr 10  & 57853.6602   &  $ 22.15$  &    0.71  &   $ -41.97$  &    2.13  &     APO \\[-4pt] 
HD 89822  &  2017 Dec 27  & 58114.7695   &  $-47.77$  &    0.68  &   $  73.27$  &    2.00  &     APO \\[-4pt] 
HD 89822  &  2018 Jan 04  & 58122.7500   &  $ 18.62$  &    0.55  &   $ -37.10$  &    1.61  &     APO \\[-4pt] 
HD 89822  &  2018 Jan 28  & 58146.9023   &  $  9.31$  &    0.50  &   $ -21.46$  &    1.47  &     APO \\[-4pt] 
HD 89822  &  2018 Apr 04  & 58212.7305   &  $ 22.88$  &    0.53  &   $ -42.29$  &    1.56  &     APO \\[-4pt] 
HD 89822  &  2019 Jan 19  & 58502.9492   &  $ 24.84$  &    0.50  &   $ -44.53$  &    1.48  &     APO \\[-4pt] 
HD 89822  &  2019 Jan 22  & 58505.7930   &  $  9.71$  &    0.53  &   $ -23.13$  &    1.57  &     APO \\[-4pt] 
HD 89822  &  2019 Jan 31  & 58514.9141   &  $ 24.61$  &    0.52  &   $ -44.97$  &    1.51  &     APO \\[-4pt] 
HD 89822  &  2019 Feb 17  & 58531.6211   &  $-48.14$  &    0.57  &   $  72.88$  &    1.66  &     APO \\[-4pt] 
HD 89822  &  2019 Mar 24  & 58566.7266   &  $-51.54$  &    0.56  &   $  78.12$  &    1.66  &     APO \\[-4pt] 
HD 89822  &  2019 Apr 21  & 58594.8203   &  $ 22.62$  &    0.22  &   $ -43.11$  &    0.49  &  Fairborn \\[-4pt] 
HD 89822  &  2019 Apr 26  & 58599.9062   &  $-18.51$  &    0.22  &   $  24.87$  &    0.49  &  Fairborn \\[-4pt] 
HD 89822  &  2019 Apr 27  & 58600.7969   &  $-42.11$  &    0.22  &   $  62.64$  &    0.49  &  Fairborn \\[-4pt] 
HD 89822  &  2019 May 02  & 58605.8711   &  $ 20.06$  &    0.22  &   $ -38.11$  &    0.49  &  Fairborn \\[-4pt] 
HD 89822  &  2019 May 04  & 58607.7891   &  $ 24.41$  &    0.22  &   $ -45.97$  &    0.49  &  Fairborn \\[-4pt] 
HD 89822  &  2019 May 18  & 58621.6367   &  $  9.95$  &    0.22  &   $ -20.78$  &    0.49  &  Fairborn \\[-4pt] 
HD 89822  &  2019 Jun 05  & 58639.7500   &  $ 13.34$  &    0.22  &   $ -27.28$  &    0.49  &  Fairborn \\[-4pt] 
HD 89822  &  2019 Jun 19  & 58653.8320   &  $ 24.66$  &    0.22  &   $ -45.41$  &    0.49  &  Fairborn \\[-4pt] 
HD 89822  &  2019 Jun 20  & 58654.7461   &  $ 22.74$  &    0.22  &   $ -43.04$  &    0.49  &  Fairborn \\[-4pt] 
HD 89822  &  2019 Jun 21  & 58655.7578   &  $ 16.49$  &    0.22  &   $ -32.00$  &    0.49  &  Fairborn \\[-4pt] 
HD 89822  &  2019 Jun 23  & 58657.7148   &  $-15.73$  &    0.22  &   $  20.75$  &    0.49  &  Fairborn \\[-4pt] 
HD 89822  &  2019 Jun 24  & 58658.7422   &  $-43.53$  &    0.22  &   $  64.73$  &    0.49  &  Fairborn \\[-4pt] 
HD 89822  &  2019 Jun 25  & 58659.7188   &  $-49.77$  &    0.22  &   $  75.39$  &    0.49  &  Fairborn \\[-4pt] 
HD 89822  &  2019 Jun 26  & 58660.6758   &  $-27.83$  &    0.22  &   $  39.81$  &    0.49  &  Fairborn \\[-4pt] 
HD 89822  &  2019 Sep 19  & 58745.9727   &  $ 24.25$  &    0.22  &   $ -45.99$  &    0.49  &  Fairborn \\[-4pt] 
HD 89822  &  2019 Oct 05  & 58761.9141   &  $-15.83$  &    0.22  &   $  19.91$  &    0.49  &  Fairborn \\[-4pt] 
HD 89822  &  2019 Oct 06  & 58762.9219   &  $-42.92$  &    0.22  &   $  63.11$  &    0.49  &  Fairborn \\[-4pt] 
HD 89822  &  2019 Oct 07  & 58763.9375   &  $-49.41$  &    0.22  &   $  73.76$  &    0.49  &  Fairborn \\[-4pt] 
HD 89822  &  2019 Oct 08  & 58764.9336   &  $-27.03$  &    0.22  &   $  38.61$  &    0.49  &  Fairborn \\[-4pt] 
HD 89822  &  2019 Oct 14  & 58770.9844   &  $ 19.99$  &    0.59  &   $ -39.24$  &    1.79  &     APO \\[-4pt] 
HD 89822  &  2019 Oct 19  & 58776.0352   &  $-39.42$  &    0.22  &   $  58.53$  &    0.49  &  Fairborn \\[-4pt] 
HD 89822  &  2019 Oct 20  & 58776.9180   &  $-16.82$  &    0.22  &   $  21.10$  &    0.49  &  Fairborn \\[-4pt] 
HD 89822  &  2019 Oct 24  & 58781.0391   &  $ 24.53$  &    0.22  &   $ -46.57$  &    0.49  &  Fairborn \\[-4pt] 
HD 89822  &  2019 Oct 25  & 58782.0391   &  $ 23.18$  &    0.22  &   $ -44.21$  &    0.49  &  Fairborn \\[-4pt] 
HD 89822  &  2019 Oct 26  & 58782.8672   &  $ 18.71$  &    0.22  &   $ -35.75$  &    0.49  &  Fairborn \\[-4pt] 
HD 89822  &  2019 Oct 30  & 58786.8555   &  $-51.43$  &    0.22  &   $  77.33$  &    0.49  &  Fairborn \\[-4pt] 
HD 89822  &  2019 Oct 31  & 58787.7812   &  $-35.30$  &    0.22  &   $  51.06$  &    0.49  &  Fairborn \\[-4pt] 
HD 89822  &  2019 Nov 14  & 58802.0273   &  $ 14.94$  &    0.54  &   $ -29.78$  &    1.57  &     APO \\[-4pt] 
HD 89822  &  2019 Nov 23  & 58810.8086   &  $-38.71$  &    0.22  &   $  56.24$  &    0.49  &  Fairborn \\[-4pt] 
HD 89822  &  2019 Dec 17  & 58834.7422   &  $-18.47$  &    0.22  &   $  23.21$  &    0.49  &  Fairborn \\[-4pt] 
HD 89822  &  2019 Dec 20  & 58837.7500   &  $ 22.06$  &    0.22  &   $ -42.07$  &    0.49  &  Fairborn \\[-4pt] 
HD 89822  &  2019 Dec 21  & 58838.7891   &  $ 24.70$  &    0.22  &   $ -46.26$  &    0.49  &  Fairborn \\[-4pt] 
HD 89822  &  2020 Jan 08  & 58856.7891   &  $-46.02$  &    0.22  &   $  68.84$  &    0.49  &  Fairborn \\[-4pt] 
HD 89822  &  2020 Jan 12  & 58860.7734   &  $ 20.97$  &    0.53  &   $ -39.66$  &    1.63  &     APO \\[-4pt] 
HD 89822  &  2020 Jan 18  & 58866.9258   &  $-37.62$  &    0.22  &   $  55.50$  &    0.49  &  Fairborn \\[-4pt] 
HD 89822  &  2020 Jan 19  & 58867.9258   &  $-51.66$  &    0.22  &   $  77.62$  &    0.49  &  Fairborn \\[-4pt] 
HD 89822  &  2020 Jan 31  & 58879.6992   &  $-49.89$  &    0.22  &   $  75.10$  &    0.49  &  Fairborn \\[-4pt] 
HD 89822  &  2020 Feb 08  & 58887.6992   &  $ 12.78$  &    0.22  &   $ -26.24$  &    0.49  &  Fairborn \\[-4pt] 
HD 89822  &  2020 Feb 15  & 58894.8906   &  $ 16.63$  &    0.22  &   $ -33.77$  &    0.49  &  Fairborn \\[-4pt] 
HD 89822  &  2020 Feb 16  & 58895.6484   &  $ 21.51$  &    0.22  &   $ -41.46$  &    0.49  &  Fairborn \\[-4pt] 
HD 89822  &  2020 Feb 17  & 58896.6367   &  $ 24.57$  &    0.22  &   $ -46.32$  &    0.49  &  Fairborn \\[-4pt] 
HD 89822  &  2020 Feb 18  & 58897.6406   &  $ 23.60$  &    0.22  &   $ -44.93$  &    0.49  &  Fairborn \\[-4pt] 
HD 89822  &  2020 Feb 19  & 58898.8320   &  $ 17.00$  &    0.22  &   $ -33.39$  &    0.49  &  Fairborn \\
\hline
HD 109510 & 2017 Feb 16   & 57800.9109 &  $  54.17$  &    0.43  &  $ -55.91$  &    1.19 &  APO       \\[-4pt]
HD 109510 & 2017 Mar 09   & 57821.7621 &  $  78.15$  &    0.46  &  $ -85.41$  &    1.26 &  APO       \\[-4pt]
HD 109510 & 2017 Dec 02   & 58089.9622 &  $ -46.53$  &    0.48  &  $  62.37$  &    1.31 &  APO       \\[-4pt]
HD 109510 & 2018 May 10   & 58248.5463 &  $  49.63$  &    0.15  &  $ -51.71$  &    1.09 &  CTIO      \\[-4pt] 
HD 109510 & 2019 Jan 13   & 58496.8454 &  $  80.35$  &    0.18  &  $ -88.34$  &    0.53 &  CTIO      \\[-4pt] 
HD 109510 & 2019 Jan 16   & 58499.8734 &  $ -22.54$  &    0.17  &  $  34.88$  &    0.49 &  CTIO      \\[-4pt] 
HD 109510 & 2019 Jan 17   & 58500.8871 &  $ -47.58$  &    0.18  &  $  63.47$  &    0.51 &  CTIO      \\[-4pt] 
HD 109510 & 2019 Jan 28   & 58511.8517 &  $  78.76$  &    0.19  &  $ -86.67$  &    0.54 &  CTIO      \\[-4pt] 
HD 109510 & 2019 Jan 29   & 58512.8259 &  $  42.81$  &    0.19  &  $ -43.54$  &    0.56 &  CTIO      \\[-4pt] 
HD 109510 & 2019 Jan 31   & 58514.9867 &  $ -34.64$  &    0.45  &  $  49.45$  &    1.18 &  APO       \\[-4pt]
HD 109510 & 2019 Feb 12   & 58526.8152 &  $  70.69$  &    0.20  &  $ -76.22$  &    0.56 &  CTIO      \\[-4pt] 
HD 109510 & 2019 Feb 13   & 58527.8282 &  $  28.73$  &    0.18  &  $ -26.07$  &    0.50 &  CTIO      \\[-4pt] 
HD 109510 & 2019 Feb 15   & 58529.8156 &  $ -38.68$  &    0.18  &  $  53.55$  &    0.49 &  CTIO      \\[-4pt] 
HD 109510 & 2019 Feb 16   & 58530.8338 &  $ -54.88$  &    0.18  &  $  72.77$  &    0.53 &  CTIO      \\[-4pt] 
HD 109510 & 2019 Feb 17   & 58531.7947 &  $ -40.67$  &    0.18  &  $  55.52$  &    0.51 &  CTIO      \\[-4pt] 
HD 109510 & 2019 Feb 18   & 58532.8140 &  $  39.66$  &    0.17  &  $ -39.42$  &    0.51 &  CTIO      \\[-4pt] 
HD 109510 & 2019 Feb 24   & 58538.8251 &  $ -50.32$  &    0.19  &  $  67.04$  &    0.55 &  CTIO      \\[-4pt] 
HD 109510 & 2019 Feb 26   & 58540.7412 &  $  77.76$  &    0.19  &  $ -85.32$  &    0.53 &  CTIO      \\[-4pt] 
HD 109510 & 2019 Feb 27   & 58541.7464 &  $  60.88$  &    0.18  &  $ -64.90$  &    0.52 &  CTIO      \\[-4pt] 
HD 109510 & 2019 Mar 05   & 58547.7628 &  $  62.38$  &    0.17  &  $ -66.44$  &    0.48 &  CTIO      \\[-4pt] 
HD 109510 & 2019 Mar 21   & 58563.7068 &  $  62.93$  &    0.10  &  $ -66.97$  &    0.77 &  CTIO      \\[-4pt] 
HD 109510 & 2019 Mar 21   & 58563.7323 &  $  62.15$  &    0.17  &  $ -66.21$  &    0.49 &  CTIO      \\[-4pt] 
HD 109510 & 2019 Mar 24   & 58566.7556 &  $ -44.78$  &    0.46  &  $  60.14$  &    1.25 &  APO       \\[-4pt]
HD 109510 & 2019 Mar 27   & 58569.7259 &  $  58.93$  &    0.18  &  $ -62.56$  &    0.49 &  CTIO      \\[-4pt] 
HD 109510 & 2019 Nov 14   & 58802.0294 &  $ -52.61$  &    0.48  &  $  69.84$  &    1.30 &  APO       \\[-4pt]
HD 109510 & 2020 Jan 07   & 58855.8570 &  $  59.03$  &    0.19  &  $ -62.26$  &    0.53 &  CTIO      \\[-4pt] 
HD 109510 & 2020 Jan 08   & 58856.8712 &  $  73.88$  &    0.20  &  $ -79.82$  &    0.56 &  CTIO      \\[-4pt] 
HD 109510 & 2020 Jan 14   & 58862.8429 &  $  28.21$  &    0.18  &  $ -25.05$  &    0.50 &  CTIO      \\[-4pt] 
HD 109510 & 2020 Jan 15   & 58863.8831 &  $  80.80$  &    0.18  &  $ -88.49$  &    0.52 &  CTIO      \\[-4pt] 
HD 109510 & 2020 Jan 16   & 58864.8731 &  $  47.75$  &    0.18  &  $ -49.18$  &    0.52 &  CTIO      \\[-4pt] 
HD 109510 & 2020 Jan 19   & 58867.8896 &  $ -50.26$  &    0.42  &  $  66.74$  &    1.25 &  CTIO      \\[-4pt] 
HD 109510 & 2020 Jan 20   & 58868.8921 &  $ -51.79$  &    0.20  &  $  68.98$  &    0.59 &  CTIO      \\[-4pt] 
HD 109510 & 2020 Jan 23   & 58871.8459 &  $  63.04$  &    0.20  &  $ -66.87$  &    0.54 &  CTIO      \\[-4pt] 
HD 109510 & 2020 Jan 26   & 58874.8406 &  $ -43.16$  &    0.44  &  $  59.32$  &    1.25 &  CTIO      \\[-4pt] 
HD 109510 & 2020 Jan 28   & 58876.8490 &  $ -28.61$  &    0.22  &  $  42.62$  &    0.64 &  CTIO      \\[-4pt] 
HD 109510 & 2020 Jan 29   & 58877.8545 &  $  58.01$  &    0.55  &  $ -61.83$  &    1.60 &  CTIO      \\[-4pt] 
HD 109510 & 2020 Jan 31   & 58879.8639 &  $  34.24$  &    0.19  &  $ -32.34$  &    0.54 &  CTIO      \\[-4pt] 
HD 109510 & 2020 Feb 14   & 58893.7607 &  $  65.90$  &    0.47  &  $ -71.83$  &    1.27 &  APO       \\[-4pt]
HD 109510 & 2020 Feb 25   & 58904.8786 &  $ -53.80$  &    0.25  &  $  72.06$  &    0.28 &  Fairborn    \\[-4pt] 
HD 109510 & 2020 May 07   & 58976.8105 &  $ -24.13$  &    0.25  &  $  36.26$  &    0.28 &  Fairborn    \\[-4pt] 
HD 109510 & 2020 Dec 21   & 59204.8936 &  $ -40.80$  &    0.25  &  $  57.10$  &    0.28 &  Fairborn    \\[-4pt] 
HD 109510 & 2021 Jan 04   & 59218.8653 &  $ -22.15$  &    0.25  &  $  34.62$  &    0.28 &  Fairborn    \\[-4pt] 
HD 109510 & 2021 Jan 14   & 59228.8270 &  $ -38.18$  &    0.25  &  $  53.64$  &    0.28 &  Fairborn    \\[-4pt] 
HD 109510 & 2021 Feb 06   & 59251.9700 &  $  53.97$  &    0.25  &  $ -56.89$  &    0.28 &  Fairborn    \\[-4pt] 
HD 109510 & 2021 Feb 23   & 59268.7125 &  $  33.74$  &    0.25  &  $ -32.85$  &    0.28 &  Fairborn    \\[-4pt] 
HD 109510 & 2021 Mar 15   & 59288.9302 &  $  72.45$  &    0.25  &  $ -78.32$  &    0.28 &  Fairborn    \\[-4pt] 
HD 109510 & 2021 Mar 20   & 59293.9990 &  $ -55.02$  &    0.25  &  $  73.54$  &    0.28 &  Fairborn    \\[-4pt] 
HD 109510 & 2021 Mar 31   & 59304.8677 &  $  56.47$  &    0.25  &  $ -60.25$  &    0.28 &  Fairborn    \\[-4pt] 
HD 109510 & 2021 Apr 10   & 59314.9131 &  $ -40.57$  &    0.25  &  $  56.27$  &    0.28 &  Fairborn    \\[-4pt] 
HD 109510 & 2021 Apr 18   & 59322.8794 &  $ -51.72$  &    0.25  &  $  69.37$  &    0.28 &  Fairborn    \\[-4pt] 
HD 109510 & 2021 Apr 30   & 59334.8455 &  $  29.79$  &    0.25  &  $ -27.46$  &    0.28 &  Fairborn    \\[-4pt] 
HD 109510 & 2021 May 04   & 59338.6809 &  $ -45.86$  &    0.25  &  $  62.51$  &    0.28 &  Fairborn    \\[-4pt] 
HD 109510 & 2021 May 05   & 59339.6808 &  $  23.70$  &    0.25  &  $ -19.79$  &    0.28 &  Fairborn    \\[-4pt] 
HD 109510 & 2021 May 06   & 59340.6808 &  $  81.24$  &    0.25  &  $ -89.27$  &    0.28 &  Fairborn    \\[-4pt] 
HD 109510 & 2021 May 07   & 59341.6807 &  $  51.46$  &    0.25  &  $ -53.23$  &    0.28 &  Fairborn    \\[-4pt] 
HD 109510 & 2021 May 09   & 59343.6806 &  $ -25.46$  &    0.25  &  $  37.79$  &    0.28 &  Fairborn    \\[-4pt] 
HD 109510 & 2021 May 10   & 59344.6804 &  $ -48.69$  &    0.25  &  $  65.92$  &    0.28 &  Fairborn    \\[-4pt] 
HD 109510 & 2021 May 11   & 59345.6804 &  $ -53.12$  &    0.25  &  $  71.12$  &    0.28 &  Fairborn    \\[-4pt] 
HD 109510 & 2021 May 13   & 59347.6803 &  $  74.44$  &    0.25  &  $ -81.92$  &    0.28 &  Fairborn    \\[-4pt] 
HD 109510 & 2021 May 14   & 59348.6802 &  $  65.22$  &    0.25  &  $ -69.70$  &    0.28 &  Fairborn    \\[-4pt] 
HD 109510 & 2021 May 15   & 59349.6801 &  $  22.51$  &    0.25  &  $ -19.40$  &    0.28 &  Fairborn    \\[-4pt] 
HD 109510 & 2021 May 16   & 59350.6800 &  $ -14.75$  &    0.25  &  $  25.09$  &    0.28 &  Fairborn    \\[-4pt] 
HD 109510 & 2021 May 17   & 59351.6800 &  $ -42.78$  &    0.25  &  $  57.94$  &    0.28 &  Fairborn    \\[-4pt] 
HD 109510 & 2021 May 18   & 59352.6799 &  $ -55.24$  &    0.25  &  $  73.34$  &    0.28 &  Fairborn    \\[-4pt] 
HD 109510 & 2021 May 19   & 59353.6798 &  $ -31.78$  &    0.25  &  $  45.14$  &    0.28 &  Fairborn    \\[-4pt] 
HD 109510 & 2021 May 24   & 59358.6794 &  $ -34.06$  &    0.25  &  $  48.42$  &    0.28 &  Fairborn    \\[-4pt] 
HD 109510 & 2021 May 25   & 59359.6793 &  $ -53.45$  &    0.25  &  $  71.30$  &    0.28 &  Fairborn    \\[-4pt] 
HD 109510 & 2021 May 26   & 59360.6793 &  $ -46.53$  &    0.25  &  $  62.81$  &    0.28 &  Fairborn    \\[-4pt] 
HD 109510 & 2021 May 28   & 59362.6791 &  $  81.34$  &    0.25  &  $ -89.00$  &    0.28 &  Fairborn    \\[-4pt] 
HD 109510 & 2021 May 29   & 59363.6790 &  $  51.51$  &    0.25  &  $ -54.07$  &    0.28 &  Fairborn    \\ 
\hline
HD 191692 &  2004 Mar 27  & 53091.9720 &  $  29.60$  &    0.89  &  $-101.50$  &    1.79   &  Fairborn    \\[-4pt]  
HD 191692 &  2004 Oct 05  & 53283.7036 &  $ -47.10$  &    1.34  &  $  -0.70$  &    2.68   &  Fairborn    \\[-4pt]  
HD 191692 &  2004 Oct 18  & 53296.6486 &  $   4.80$  &    0.89  &  $ -74.40$  &    1.79   &  Fairborn    \\[-4pt]  
HD 191692 &  2004 Oct 24  & 53302.7319 &  $ -53.60$  &    1.34  &  $   3.70$  &    2.68   &  Fairborn    \\[-4pt]  
HD 191692 &  2004 Oct 25  & 53303.6305 &  $ -53.70$  &    1.34  &  $   3.70$  &    2.68   &  Fairborn    \\[-4pt]  
HD 191692 &  2004 Nov 05  & 53314.6562 &  $  28.80$  &    0.89  &  $-101.90$  &    1.79   &  Fairborn    \\[-4pt]  
HD 191692 &  2004 Nov 11  & 53320.6595 &  $ -52.60$  &    1.34  &  $   4.50$  &    2.68   &  Fairborn    \\[-4pt]  
HD 191692 &  2014 Jul 03  & 56841.8663 &  $  17.37$  &    0.24  &  $ -90.68$  &    0.36   &  CTIO    	\\[-4pt]  
HD 191692 &  2019 Jun 06  & 58640.8262 &  $  43.15$  &    0.32  &  $-123.37$  &    0.46   &  CTIO    	\\[-4pt]  
HD 191692 &  2019 Aug 12  & 58707.6604 &  $   0.24$  &    0.25  &  $ -68.29$  &    0.38   &  CTIO    	\\[-4pt]  
HD 191692 &  2019 Aug 13  & 58708.6722 &  $  25.08$  &    0.30  &  $-100.51$  &    0.45   &  CTIO    	\\[-4pt]  
HD 191692 &  2019 Aug 14  & 58709.6792 &  $  39.63$  &    0.36  &  $-119.35$  &    0.51   &  CTIO    	\\[-4pt]  
HD 191692 &  2019 Aug 17  & 58712.6434 &  $ -51.92$  &    0.45  &  $   0.28$  &    0.60   &  CTIO    	\\[-4pt]  
HD 191692 &  2019 Aug 29  & 58724.6105 &  $  -2.38$  &    0.25  &  $ -64.37$  &    0.37   &  CTIO    	\\[-4pt]  
HD 191692 &  2019 Aug 30  & 58725.6912 &  $  21.91$  &    0.28  &  $ -96.19$  &    0.41   &  CTIO    	\\[-4pt]  
HD 191692 &  2019 Aug 31  & 58726.5889 &  $  43.76$  &    0.27  &  $-124.37$  &    0.39   &  CTIO    	\\[-4pt]  
HD 191692 &  2019 Sep 03  & 58729.6022 &  $ -50.62$  &    0.42  &  $  -0.36$  &    0.54   &  CTIO    	\\[-4pt]  
HD 191692 &  2019 Sep 04  & 58730.5631 &  $ -53.58$  &    0.42  &  $   2.39$  &    0.57   &  CTIO    	\\[-4pt]  
HD 191692 &  2019 Sep 05  & 58731.5943 &  $ -53.72$  &    0.46  &  $   2.28$  &    0.63   &  CTIO    	\\[-4pt]  
HD 191692 &  2019 Sep 15  & 58741.5586 &  $  -5.50$  &    0.27  &  $ -60.96$  &    0.37   &  CTIO    	\\[-4pt]  
HD 191692 &  2019 Sep 16  & 58742.5689 &  $  14.65$  &    0.26  &  $ -87.40$  &    0.38   &  CTIO    	\\[-4pt]  
HD 191692 &  2019 Sep 17  & 58743.5649 &  $  42.85$  &    0.33  &  $-123.43$  &    0.47   &  CTIO    	\\[-4pt]  
HD 191692 &  2019 Oct 05  & 58761.6219 &  $   3.61$  &    0.28  &  $ -72.76$  &    0.41   &  CTIO    	\\[-4pt]  
HD 191692 &  2019 Oct 10  & 58766.5178 &  $ -52.65$  &    0.45  &  $   0.97$  &    0.62   &  CTIO    	\\[-4pt]  
HD 191692 &  2019 Oct 10  & 58766.6965 &  $ -54.20$  &    1.34  &  $  -1.20$  &    2.68   &  Fairborn    \\[-4pt]  
HD 191692 &  2019 Oct 10  & 58767.4814 &  $ -50.46$  &    0.61  &  $  -0.37$  &    0.78   &  CTIO    	\\[-4pt]  
HD 191692 &  2019 Oct 11  & 58767.5990 &  $ -52.20$  &    1.34  &  $  -0.60$  &    2.68   &  Fairborn    \\[-4pt]  
HD 191692 &  2019 Oct 19  & 58775.6700 &  $  -6.00$  &    0.89  &  $ -57.60$  &    1.79   &  Fairborn    \\[-4pt]  
HD 191692 &  2019 Oct 20  & 58776.5546 &  $   8.29$  &    0.47  &  $ -79.32$  &    0.70   &  CTIO    	\\[-4pt]  
HD 191692 &  2019 Oct 20  & 58776.6822 &  $  12.50$  &    0.89  &  $ -82.40$  &    1.79   &  Fairborn    \\[-4pt]  
HD 191692 &  2019 Oct 21  & 58777.5640 &  $  37.04$  &    0.32  &  $-116.10$  &    0.47   &  CTIO    	\\[-4pt]  
HD 191692 &  2019 Oct 21  & 58777.6927 &  $  39.70$  &    0.89  &  $-121.60$  &    1.79   &  Fairborn    \\[-4pt]  
HD 191692 &  2019 Oct 22  & 58778.5228 &  $  19.41$  &    0.31  &  $ -92.89$  &    0.44   &  CTIO    	\\[-4pt]  
HD 191692 &  2019 Oct 22  & 58778.6162 &  $  11.90$  &    0.89  &  $ -83.00$  &    1.79   &  Fairborn    \\[-4pt]  
HD 191692 &  2019 Oct 25  & 58781.5327 &  $ -52.55$  &    0.44  &  $   1.49$  &    0.60   &  CTIO    	\\[-4pt]  
HD 191692 &  2019 Oct 26  & 58782.6390 &  $ -54.40$  &    1.34  &  $   1.70$  &    2.68   &  Fairborn    \\[-4pt]  
HD 191692 &  2019 Oct 27  & 58783.6384 &  $ -53.70$  &    1.34  &  $   2.50$  &    2.68   &  Fairborn    \\[-4pt]  
HD 191692 &  2019 Nov 06  & 58793.5064 &  $   4.57$  &    0.29  &  $ -74.38$  &    0.44   &  CTIO    	\\[-4pt]  
HD 191692 &  2019 Nov 07  & 58795.4940 &  $  29.78$  &    0.31  &  $-106.24$  &    0.45   &  CTIO    	\\[-4pt]  
HD 191692 &  2020 Sep 11  & 59103.6197 &  $  36.50$  &    0.89  &  $-115.20$  &    1.79   &  Fairborn    \\[-4pt]  
HD 191692 &  2020 Sep 16  & 59108.6166 &  $ -56.10$  &    1.34  &  $   4.50$  &    2.68   &  Fairborn    \\[-4pt]  
HD 191692 &  2020 Sep 17  & 59109.6126 &  $ -52.20$  &    1.34  &  $   2.70$  &    2.68   &  Fairborn    \\[-4pt]  
HD 191692 &  2020 Sep 25  & 59117.6385 &  $ -13.30$  &    0.89  &  $ -50.00$  &    1.79   &  Fairborn    \\[-4pt]  
HD 191692 &  2020 Sep 26  & 59118.6016 &  $   1.50$  &    0.89  &  $ -67.30$  &    1.79   &  Fairborn    \\[-4pt]  
HD 191692 &  2020 Sep 27  & 59119.6038 &  $  25.20$  &    0.89  &  $ -95.50$  &    1.79   &  Fairborn    \\[-4pt]  
HD 191692 &  2020 Sep 28  & 59120.6023 &  $  41.70$  &    0.89  &  $-125.10$  &    1.79   &  Fairborn    \\[-4pt]  
HD 191692 &  2020 Sep 29  & 59121.6015 &  $ -17.60$  &    0.89  &  $ -44.60$  &    1.79   &  Fairborn    \\[-4pt]  
HD 191692 &  2020 Sep 30  & 59122.6014 &  $ -44.50$  &    1.34  &  $  -8.00$  &    2.68   &  Fairborn    \\[-4pt]  
HD 191692 &  2020 Oct 01  & 59123.6002 &  $ -53.10$  &    1.34  &  $  -0.20$  &    2.68   &  Fairborn    \\[-4pt]  
HD 191692 &  2020 Oct 02  & 59124.5996 &  $ -54.30$  &    1.34  &  $   3.60$  &    2.68   &  Fairborn    \\[-4pt]  
HD 191692 &  2020 Oct 03  & 59125.5985 &  $ -54.80$  &    1.34  &  $   3.10$  &    2.68   &  Fairborn    \\[-4pt]  
HD 191692 &  2020 Oct 04  & 59126.5986 &  $ -53.50$  &    1.34  &  $   2.10$  &    2.68   &  Fairborn    \\[-4pt]  
HD 191692 &  2020 Oct 05  & 59127.5947 &  $ -50.10$  &    1.34  &  $  -1.20$  &    2.68   &  Fairborn    \\[-4pt]  
HD 191692 &  2020 Oct 06  & 59128.5935 &  $ -47.90$  &    1.34  &  $  -4.30$  &    2.68   &  Fairborn    \\[-4pt]  
HD 191692 &  2020 Oct 13  & 59135.6121 &  $  -1.70$  &    0.89  &  $ -64.30$  &    1.79   &  Fairborn    \\[-4pt]  
HD 191692 &  2020 Oct 14  & 59136.5880 &  $  19.80$  &    0.89  &  $ -90.60$  &    1.79   &  Fairborn    \\[-4pt]  
HD 191692 &  2020 Oct 15  & 59137.5874 &  $  44.40$  &    0.89  &  $-127.00$  &    1.79   &  Fairborn    \\[-4pt]  
HD 191692 &  2020 Oct 16  & 59138.5866 &  $  -9.80$  &    0.89  &  $ -52.00$  &    1.79   &  Fairborn    \\[-4pt]  
HD 191692 &  2020 Oct 18  & 59140.5850 &  $ -52.90$  &    1.34  &  $   0.80$  &    2.68   &  Fairborn    \\[-4pt]  
HD 191692 &  2020 Oct 19  & 59141.5841 &  $ -52.90$  &    1.34  &  $   3.40$  &    2.68   &  Fairborn    \\[-4pt]  
HD 191692 &  2020 Oct 20  & 59142.6096 &  $ -53.10$  &    1.34  &  $   3.40$  &    2.68   &  Fairborn    \\[-4pt]  
HD 191692 &  2020 Oct 21  & 59143.6136 &  $ -53.60$  &    1.34  &  $   2.30$  &    2.68   &  Fairborn    \\[-4pt]  
HD 191692 &  2020 Oct 22  & 59144.6130 &  $ -50.60$  &    1.34  &  $  -1.00$  &    2.68   &  Fairborn    \\[-4pt]  
HD 191692 &  2020 Oct 30  & 59152.6153 &  $  -2.70$  &    0.89  &  $ -61.10$  &    1.79   &  Fairborn    \\[-4pt]  
HD 191692 &  2020 Oct 31  & 59153.6148 &  $  17.40$  &    0.89  &  $ -88.10$  &    1.79   &  Fairborn    \\[-4pt]  
HD 191692 &  2020 Nov 01  & 59154.6248 &  $  44.20$  &    0.89  &  $-125.90$  &    1.79   &  Fairborn    \\[-4pt]  
HD 191692 &  2020 Nov 02  & 59155.6256 &  $  -4.50$  &    0.89  &  $ -58.50$  &    1.79   &  Fairborn    \\[-4pt]  
HD 191692 &  2020 Nov 04  & 59157.6228 &  $ -52.50$  &    1.34  &  $   0.50$  &    2.68   &  Fairborn    \\[-4pt]  
HD 191692 &  2020 Nov 05  & 59158.6216 &  $ -53.80$  &    1.34  &  $   3.10$  &    2.68   &  Fairborn    \\[-4pt]  
HD 191692 &  2020 Nov 17  & 59170.6104 &  $  13.40$  &    0.89  &  $ -83.00$  &    1.79   &  Fairborn    \\[-4pt]  
HD 191692 &  2020 Nov 18  & 59171.6560 &  $  43.50$  &    0.89  &  $-127.20$  &    1.79   &  Fairborn    \\[-4pt]  
HD 191692 &  2020 Nov 19  & 59172.6341 &  $   4.00$  &    0.89  &  $ -69.90$  &    1.79   &  Fairborn    \\[-4pt]  
HD 191692 &  2020 Nov 21  & 59174.6069 &  $ -51.50$  &    1.34  &  $  -1.90$  &    2.68   &  Fairborn    \\[-4pt]  
HD 191692 &  2020 Nov 22  & 59175.6065 &  $ -54.00$  &    1.34  &  $   5.20$  &    2.68   &  Fairborn    \\[-4pt]  
HD 191692 &  2020 Nov 25  & 59178.6170 &  $ -51.50$  &    1.34  &  $   0.70$  &    2.68   &  Fairborn    \\[-4pt]  
HD 191692 &  2021 Feb 12  & 59258.0383 &  $  17.70$  &    0.89  &  $ -92.90$  &    1.79   &  Fairborn    \\[-4pt]  
HD 191692 &  2021 Feb 28  & 59274.0231 &  $  32.70$  &    0.89  &  $-109.50$  &    1.79   &  Fairborn    \\[-4pt]  
HD 191692 &  2021 Mar 01  & 59275.0178 &  $  26.70$  &    0.89  &  $-102.80$  &    1.79   &  Fairborn    \\[-4pt]  
HD 191692 &  2021 Apr 09  & 59314.0086 &  $ -54.40$  &    1.34  &  $   2.40$  &    2.68   &  Fairborn    \\[-4pt]  
HD 191692 &  2021 Apr 20  & 59324.8890 &  $  18.10$  &    0.89  &  $ -90.00$  &    1.79   &  Fairborn    \\[-4pt]  
HD 191692 &  2021 Apr 24  & 59328.8698 &  $ -52.20$  &    1.34  &  $   0.20$  &    2.68   &  Fairborn    \\[-4pt]  
HD 191692 &  2021 May 07  & 59341.8248 &  $  12.30$  &    0.89  &  $ -82.30$  &    1.79   &  Fairborn    \\[-4pt]  
HD 191692 &  2021 Jun 01  & 59366.8105 &  $ -52.70$  &    1.34  &  $  -1.20$  &    2.68   &  Fairborn    \\
\enddata
\end{deluxetable*}

%% file: table_cal.txt
\begin{deluxetable*}{lccc}
\tablewidth{0pt}
\tabletypesize{\small}
\tablecaption{Calibrator Angular Diameters \label{cal}}
\tablehead{
\colhead{Target} &\colhead{Calibrator} & \colhead{$\theta_{\rm UD}$ (mas)} & \colhead{Wavelength}
}
\startdata	            
HD 61859 &  HD 45391    & $0.329\pm0.017$   & $H$-band  \\ 
HD 61859 &  HD 56124    & $0.346\pm0.008$   & $K$-band  \\ 
HD 61859 &  HD 59037    & $0.391\pm0.011$   & $K$-band  \\ 
HD 61859 &  HD 59747    & $0.348\pm0.017$   & $K$-band  \\ 
HD 61859 &  HD 63495    & $0.122\pm0.006$   & $K$-band  \\ 
HD 61859 &  HD 67709    & $0.443\pm0.010$   & $K$-band  \\ 
HD 61859 &  HD 67827    & $0.387\pm0.019$   & $K$-band  \\ 
HD 61859 &  HD 72524    & $0.256\pm0.013$   & $H$-band  \\ 
HD 89822 &  HD 84812    & $0.300\pm0.015$   & $H$-band  \\ 
HD 89822 &  HD 88983    & $0.313\pm0.008$   & $K$-band  \\ 
HD 89822 &  HD 96707    & $0.296\pm0.007$   & $K$-band  \\ 
HD 89822 &  HD 98772    & $0.235\pm0.012$   & $H$-band  \\ 
HD 109510 &  HD 105086   & $0.319\pm0.016$   & $H$-band  \\ 
HD 109510 &  HD 107569   & $0.242\pm0.006$   & $K$-band  \\ 
HD 109510 &  HD 111718   & $0.231\pm0.005$   & $K$-band  \\ 
HD 109510 &  HD 111893   & $0.263\pm0.013$   & $H$-band  \\ 
HD 191692 &  HD 188350   & $0.284\pm0.008$   & $K$-band  \\ 
HD 191692 &  HD 191014   & $0.826\pm0.073$   & $K$-band  \\ 
HD 191692 &  HD 193329   & $0.831\pm0.082$   & $K$-band  \\ 
HD 191692 &  HD 195810   & $0.323\pm0.029$   & $K$-band  \\ 
HD 191692 &  HD 185124   & $0.495\pm0.025$   & $H$-band  \\ 
HD 191692 &  HD 196870   & $0.661\pm0.050$   & $H$-band  \\ 
\enddata     
\end{deluxetable*}